\documentclass[twocolumn,aps,showpacs,floats,pre]{revtex4-1}
\usepackage[usenames,dvipsnames]{color}
\usepackage{amsmath}
\usepackage{amssymb}
\usepackage{graphicx}
\usepackage{anysize}
\usepackage{mathrsfs}
\usepackage{notes2bib}
\usepackage{multirow}
\usepackage{relsize}
\usepackage[final]{pdfpages}

\marginsize{1.5cm}{1.5cm}{0.5cm}{1.cm}
 
\usepackage{breakurl}

\allowdisplaybreaks
\makeatletter
\@ifundefined{textcolor}{}
{
 \definecolor{BLACK}{gray}{0}
 \definecolor{WHITE}{gray}{1}
 \definecolor{RED}{rgb}{1,0,0}
 \definecolor{GREEN}{rgb}{0,1,0}
 \definecolor{BLUE}{rgb}{0,0,1}
 \definecolor{CYAN}{cmyk}{1,0,0,0}
 \definecolor{MAGENTA}{cmyk}{0,1,0,0}
 \definecolor{YELLOW}{cmyk}{0,0,1,0}
 \definecolor{auburn}{rgb}{0.43, 0.21, 0.1}
}

\usepackage[unicode=true,pdfusetitle, bookmarks=true,bookmarksnumbered=false,bookmarksopen=false, breaklinks=false,pdfborder={0 0 1},backref=false,colorlinks=true]{hyperref}
\hypersetup{colorlinks=true,
	linkcolor=black,
	anchorcolor=black,
	citecolor=black,
	urlcolor=black
}

\usepackage{graphics}
\usepackage{epsfig}
\usepackage{epsf}

\providecommand{\de}{\delta}
\providecommand{\tn}{\tabularnewline}
\providecommand{\nn}{\nonumber}
\providecommand{\be}{\begin{equation}}
\providecommand{\ee}{\end{equation}}
\providecommand{\bea}{\begin{eqnarray}}
\providecommand{\eea}{\end{eqnarray}}
\providecommand{\beas}{\begin{eqnarray*}}
\providecommand{\eeas}{\end{eqnarray*}}

\providecommand{\beni}{\begin{equation*}}
\providecommand{\eeni}{\end{equation*}}

\providecommand{\bw}{\begin{widetext}}
\providecommand{\ew}{\end{widetext}}
\providecommand{\imi}{\mathrm{i}}

\makeatletter
\newcommand{\vast}{\bBigg@{3}}
\newcommand{\Vast}{\bBigg@{4}}
\makeatother

\makeatother

\begin{document}

\title{Hierarchical spin glasses in a magnetic field: A renormalization-group study}
\author{Michele Castellana}
\affiliation{Joseph Henry Laboratories of Physics and Lewis--Sigler Institute for Integrative Genomics, Princeton University, Princeton, New Jersey 08544, United States}
\author{Carlo Barbieri}
\affiliation{Dipartimento di Fisica, Sapienza Universit\`{a} di Roma, Piazzale Aldo Moro 5, 00185, Rome, Italy}


\begin{abstract}
By using renormalization-group (RG) methods, we study a non-mean-field model of a spin glass built on a hierarchical lattice, the hierarchical Edwards-Anderson model in a magnetic field. We investigate the spin-glass transition in a field by studying the existence of a stable critical RG fixed point (FP) with perturbation theory. In the parameter region where the model has a mean-field behavior---corresponding to $d \geq 4$ for a $d$-dimensional Ising model---we find a stable  FP that corresponds to a spin-glass transition in a field. In the non-mean-field parameter region the  FP above is unstable, and we determined exactly all other FPs: to our knowledge, this is the first time that all perturbative FPs for the full set of RG equations of a spin glass in a field have been characterized in the non-mean-field region. We find that all potentially stable FPs in the non-mean-field region have a nonzero imaginary part: this constitutes, to the best of our knowledge, the first demonstration for a spin glass in a field that there is no perturbative FP corresponding to a spin-glass transition in the non-mean-field region. Finally, we discuss the possible interpretations of this result, such as the absence of a phase transition in a field, or the existence of a transition associated with a non-perturbative FP. 
\end{abstract}

\maketitle

\section{Introduction}

Establishing the existence of a phase transition in non-mean-field spin glasses in a magnetic field is an open problem that has been attracting growing interest in the last few decades. Indeed, a demonstration of the existence or of the absence of such a transition could shed light on a fundamental problem: understanding the structure of the low-temperature phase of spin glasses. Namely, the occurrence of a transition in a field would indicate that the energy landscape of a spin glass has a complex structure involving an exponentially large number of states \cite{mezard1987spin}, while the absence of this transition would hint the existence of only two low-lying energy basins, which are reminiscent of the ground states of Ising ferromagnets \cite{fisher1986ordered}. 
Despite decades of intense research, the occurrence of a phase transition for non-mean-field spin glasses in a magnetic field  is still under debate: Indeed, experimental studies on disordered magnetic materials provided evidence both in favor \cite{petit1999ordering} and against \cite{jonsson2005dynamical}  a transition. In addition, numerical simulations for Ising spin glasses with short and long-range interactions are affected by long equilibration times and strong finite-size effects, preventing numerical approaches from giving a definite answer on the existence of a transition in a field \cite{jorg2008behavior,young2004absence,banos2012thermodynamic,katzgraber2009study,leuzzi2009ising}.

On a conceptual level, one of the reasons why non-mean-field models of spin glasses are hard to solve is that the complex structure of the low-lying energy states hinders the use of the renormalization-group (RG) coarse-graining methods widely employed for  homogeneous systems, such as Ising ferromagnets \cite{wilson1974renormalization}.   
In this regard, the RG transformation for ferromagnetic spin systems  takes a strikingly simple form when applied to models built on a hierarchical lattice \cite{dyson1969existence}. 
To study non-mean-field spin glasses with RG methods, it is thus natural to consider a spin-glass model where spin interactions are disposed in a hierarchical way. This model is known as the hierarchical Edwards-Anderson model \cite{franz2009overlap}, and it has been raising interest in recent years because it provides a natural implementation of the RG transformation, allowing for a novel RG characterization of the critical properties of a non-mean-field spin glass \cite{castellana2011renormalization,castellana2010renormalization,castellana2011real,angelini2013ensemble}.   

In this paper, we study the existence of a phase transition for the hierarchical Edwards-Anderson model with an external magnetic field (HEAM). The hierarchical structure of spin-spin interactions for the HEAM implies an exact RG equation for the replicated partition function with fixed overlap which is identical to the RG equation for the hierarchical Edwards-Anderson model \cite{franz2009overlap}. In perturbation theory, this RG equation takes a particularly involved form: indeed, the presence of the external field implies that the replicated partition function is given by a combination of twelve overlap monomials \cite{pimentel2002spin}, resulting into a complex perturbative structure. To carry out such an involved perturbative expansion, we developed a symbolic manipulation tool which allowed us for deriving the RG recurrence equations to lowest order in perturbation theory. We then investigated the existence of a phase transition in a field by studying the  fixed points (FPs) of these RG equations. The FP equations are a system of polynomial equations of the third degree which may have, in principle, a large number of solutions. 
In the parameter region where the HEAM has a mean-field behavior we find a stable FP corresponding to the existence of a spin-glass transition. In the non-mean-field region the FP above is unstable, and all other solutions of the FP equations need to be studied to  establish the existence of a stable FP. 
In this regard, to our knowledge the full set of perturbative FPs in the non-mean-field region of a spin glass in a field has never been fully characterized: in particular, two previous RG studies addressed this problem for short-range spin glasses in a field. Pimentel  \textit{et al.} found a set of unstable FPs \cite{pimentel2002spin}, but these were not shown to coincide with the full set of solutions of the FP equations \cite{pimentel2014private}: as a consequence, the possibility that other stable FPs could exist remained open. Other RG approaches considered only certain linear combinations of the overlap matrix---the `replicon' modes---thus retaining only a subset of the twelve monomials in the replicated partition function \cite{bray1980renormalisation,moore2011disappearance}: this approach resulted in a reduced set of FP equations which were found to have no stable FP.
By using the Gr\"{o}bner-basis method for systems of polynomial equations \cite{froberg1997introduction}, here we determined  exactly all perturbative FPs of the full set of RG equations in the non-mean-field region of  the HEAM, and we show that all potentially stable FPs have a nonzero imaginary part. To our knowledge, this result constitutes an unprecedented demonstration that there is no perturbative FP corresponding to a spin-glass transition in the non-mean-field region of a spin glass in a field.

The paper is organized as follows: In Section \ref{sec1} we introduce the HEAM, in Section \ref{sec1.1} we consider the RG equations and we discuss their perturbative solution, in Section \ref{sec1.2} we derive the qualitative structure of the critical FP, which is then determined explicitly in Section \ref{sec1.3}. Finally, Section \ref{sec2} is devoted to the discussion and interpretation of the results, as well as to an outlook of future studies.

\begin{figure}
\centering\includegraphics[scale=0.45]{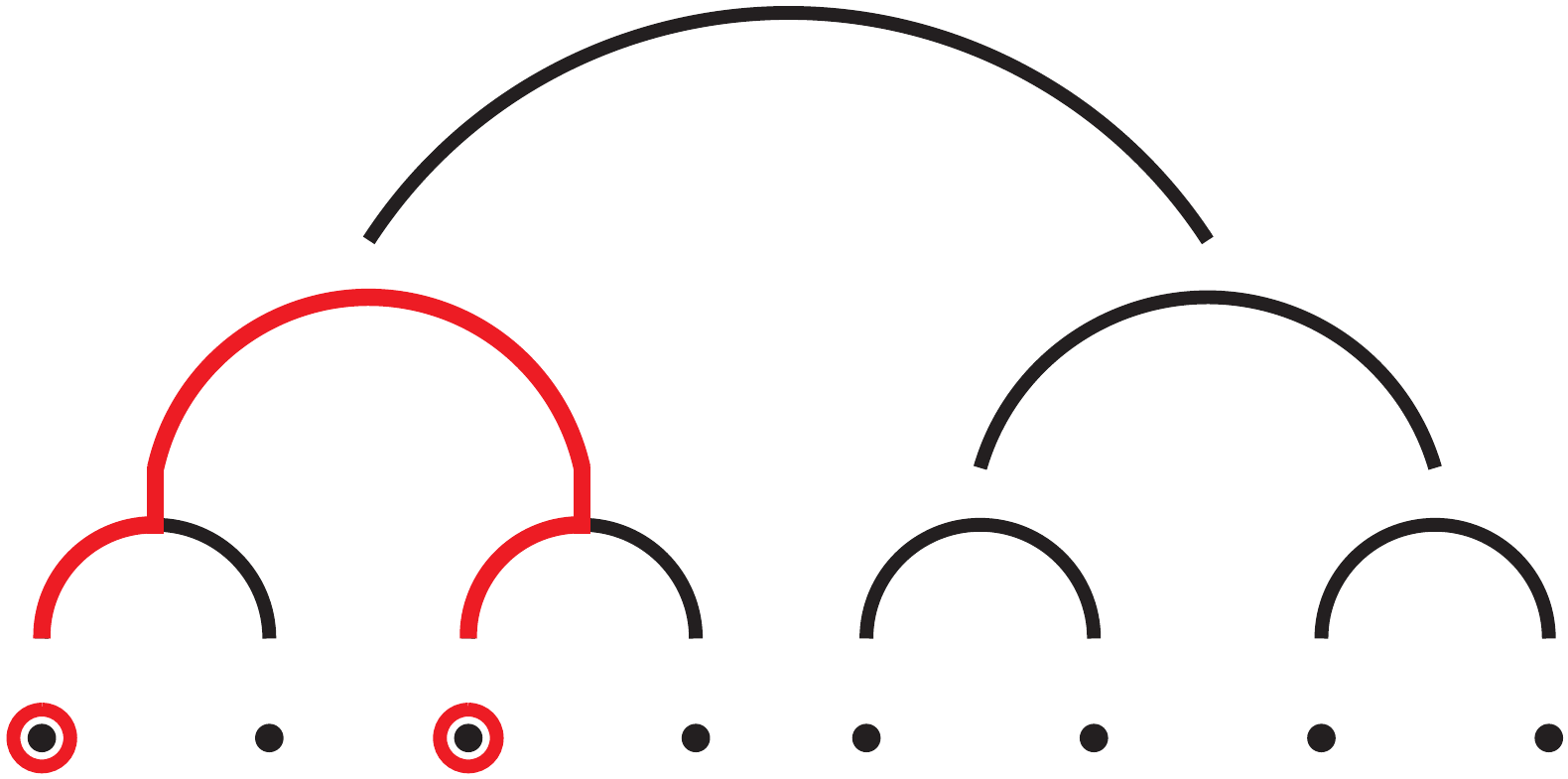}
\caption{
Hierarchical Edwards-Anderson model in a magnetic field with $k=3$.  Dots represents spins $S_1, \ldots, S_8$ from left to right,  arcs represent  interactions between spins below them, and magnetic fields are not shown. The red path starts from the two circled  spins $S_1$, $S_3$ and it goes from the bottom to the top of the hierarchical tree until a common arc between the two spins is found: since this requires ascending two hierarchical levels, the hierarchical distance between $S_1$ and $S_3$ is $d_{13} = 2$. \label{fig1}}
\end{figure}

\section{Results}\label{sec1}

The HEAM is a system of $2^k$ Ising spins $\vec{S} \equiv S_1, \ldots, S_{2^k}$, $S_i = \pm 1$, whose Hamiltonian $H_k[\vec{S}]$ is given by the recursive equation \cite{franz2009overlap}
\be\label{ghea}
H_k[\vec{S}]=H_{k-1}^{1}[\vec{S}_{1}]+H_{k-1}^{2}[\vec{S}_{2}]-\frac{1}{2^{k\sigma}}\sum_{i<j=1}^{2^k}J_{ij}S_{i}S_{j},
\ee 
where $\vec{S}_1 \equiv S_1, \ldots, S_{2^{k-1}}$ and $\vec{S}_2 \equiv S_{2^{k-1}+1}, \ldots, S_{2^k}$ denote the spins in the left and right half of the lattice respectively,   $1/2 < \sigma < 1$ is a parameter which determines how fast spin-spin interactions decrease with distance \bibnote[note1]{Here we chose $1/2 < \sigma < 1$ because this is the interval where the hierarchical Edwards-Anderson model is expected to have a finite-temperature phase transition \cite{franz2009overlap}.}, $\{ J_{ij} \}$ are independent and identically distributed (IID) Gaussian random variables with zero mean and unit variance, and the single-spin Hamiltonian on site $i$ is $H_0[S_i] = h_i S_i$, where $\{ h_i \}$ are IID Gaussian random variables with zero mean. 
Given two spins $S_i$, $S_j$, let us consider the number of levels that we need to ascend in the hierarchical tree starting from the bottom in order to find a root common to $S_i$ and $S_j$, see Fig. \ref{fig1}: we denote this number by the hierarchical distance $d_{ij}$. The recursive structure of the Hamiltonian (\ref{ghea}) then implies that the interaction strength between spins $S_i$, $S_j$ depends on $i$, $j$ through their hierarchical distance rather than through their Euclidean distance: indeed, Eq. (\ref{ghea}) can be rewritten as $H_k[\vec{S}]=-\sum_{i<j=1}^{2^k}K_{ij}S_{i}S_{j} + \sum_{i=1}^{2^k} h_i S_i$, where $\{ K_{ij} \}$ are IID Gaussian random variables, and $K_{ij}$ has zero mean and variance $\big( \sum_{l=0}^{k-d_{ij}} 2^{-2 \sigma l} \big) 2^{-2 \sigma d_{ij}}$.

To relate the HEAM to other spin-glass models with different disorder distributions such as short- and long-range Ising spin glasses \cite{edwards1975theory,kotliar1983one}, two considerations are in order. First, a precise correspondence between  short-range spin models in $d$ dimensions and one-dimensional long-range models is still subject of  research  \cite{leuzzi2013long,katzgraber2009study}. Second, the universality of physical observables with respect to the form of the bond distribution in Ising spin glasses is also a subject of ongoing debate  \cite{bernardi1995violation,lundow2013critical,lundow2014ising,lundow2014evidence}.
Despite the considerations above, a hierarchical Edwards-Anderson model  with a given value of $\sigma$ has some features in common with a short-range Ising spin glass on a hypercubic $d$-dimensional lattice \cite{castellana2011renormalization}. In particular, the parameter region $1/2 < \sigma \leq 2/3$, where the hierarchical Edwards-Anderson model has a mean-field behavior, is analogous to $d \geq 6$ for a short-range Ising spin glass, i.e. it is characterized by vanishing order-parameter fluctuations. Similarly, the non-mean-field region $2/3 < \sigma <1$ corresponds to $d<6$ for an Ising spin glass, and it is characterized by a non-Gaussian RG FP associated with nonzero order-parameter fluctuations.

\subsection{Renormalization-group equations}\label{sec1.1}

The hierarchical structure of the interactions in the HEAM shown in  Fig. \ref{fig1} implies an exact RG equation which can be derived with the replica method \cite{castellana2011renormalization}: we introduce an integer $n$,  an $n \times n$  matrix $Q_{ab}$ with $Q_{ab} = Q_{ba}$ and $Q_{aa} = 0$, and the replicated partition function with fixed overlap
\be\label{eq2}
\hspace{-1.5mm}Z_k[Q] \equiv  \mathbb{E}\vast[ \sum_{\{ \vec{S}^a \}} e^{ - \mathlarger{\beta \sum_{a=1}^n H_k[\vec{S}^a]}} \hspace{-1mm} \prod_{a<b=1}^n \hspace{-2mm} \de ( Q_{ab} - q_{ab} ) \vast],
\ee
where in Eq. (\ref{eq2}) $\mathbb{E}[ \, ]$ denotes the expectation with respect to all random variables, $\beta = 1/T$ is the inverse temperature,  $\{ \vec{S}^a \}$ are $n$ replicas of the spin configuration $\vec{S}$, and 
\be
q_{ab} \equiv \frac{1}{2^k}\sum_{i=1}^{2^k} S^a_i S^b_i
\ee is the overlap \cite{mezard1987spin} between replicas $a$ and $b$. As we will show in what follows, the replicated partition function  (\ref{eq2}) is a useful quantity because, in principle, its $n \rightarrow 0$ limit incorporates all thermodynamical properties of the system \cite{mezard1987spin}. Let us introduce
\be\label{eq21}
C \equiv 2^{2(1-\sigma)}
\ee
and the rescaled overlap distribution
\be\label{eq13}
\mathcal{Z}_k[Q] \equiv Z_k[C^{-k/2}Q].
\ee
By using Eqs. (\ref{ghea}), (\ref{eq2}), it  can be shown \cite{franz2009overlap,castellana2011renormalization} that $\mathcal{Z}_k$ satisfies the following recursive equation
\bw
\be\label{eq3}
\mathcal{Z}_{k+1}[Q] = \exp \left(  \frac{\beta^2}{4} \text{Tr}[Q^2] \right)  \int [d P]  \mathcal{Z}_k\left[\frac{Q+P}{C^{1/2}}\right]\mathcal{Z}_k\left[\frac{Q-P}{C^{1/2}}\right],
\ee
\ew
where $\int [d P] \equiv \int \prod_{a<b} dP_{ab}$, and in what follows we will omit $Q$-independent proportionality factors multiplying $\mathcal{Z}_k$.

\begin{centering}
\begin{table}
\begin{centering}
\begin{tabular}{|l|l|l|}
\hline 
$\mathrm{I}^1[Q]$ & $\mathrm{I}^2_p[Q]$ & $\mathrm{I}^3_p[Q]$\tn
\hline
$\sum_{a,b=1}^n Q_{ab}$ & $\sum_{a,b=1}^n Q_{ab}^2$ & $\sum_{a,b,c=1}^n Q_{ab}Q_{bc}Q_{ca}$ \tn
 & $\sum_{a,b,c=1}^n Q_{ab} Q_{ac}$ & $\sum_{a,b=1}^n Q_{ab}^3$ \tn
 & $\sum_{a,b,c,d=1}^n Q_{ab}Q_{cd}$ & $\sum_{a,b,c=1}^n Q_{ab}^2Q_{ac}$ \tn
 & & $\sum_{a,b,c,d=1}^n Q_{ab}^2 Q_{cd}$ \tn
 & & $\sum_{a,b,c,d=1}^nQ_{ab}Q_{ac}Q_{bd}$ \tn
 & & $\sum_{a,b,c,d=1}^nQ_{ab}Q_{ac}Q_{ad}$ \tn
& & $\sum_{a,b,c,d,e=1}^n Q_{ab}Q_{ac} Q_{de}$ \tn
& & $\sum_{a,b,c,d,e,f=1}^n Q_{ab}Q_{cd}Q_{ef}$ \tn
\hline
\end{tabular}
\par\end{centering}

\caption{Monomials contributing to the overlap probability distribution of the  hierarchical Edwards-Anderson model in a magnetic field to order $Q^3$. The monomials of order one, two and three are listed in the rows of the first, second and third column respectively, from top to bottom in increasing order of the index $p$.\label{tab1}}
\end{table}
\end{centering}

Being a functional recursive equation, Eq. (\ref{eq3}) is hard to solve with exact methods: still, an analytical solution can be worked out in perturbation theory. We write $\mathcal{Z}_k[Q]$ as a combination of  monomials in $Q$, where each monomial is multiplied by a numerical coefficient which depends on $k$: then, Eq. (\ref{eq3}) implies a set of algebraic recursive equations for these coefficients, and we will show that these equations can be studied analytically in perturbation theory. Neglecting monomials of order larger than $Q^3$, the most general form of $\mathcal{Z}_k[Q]$ is given by \cite{temesvari2008almeida}
\bea\label{eq4}\nn
\mathcal{Z}_k[Q]  & = &\\
\exp\left[-\left( s_k \, \mathrm{I}^1[Q] + \frac{1}{2} \sum_{p=1}^3 r^p_k \, \mathrm{I}^2_p[Q] + \frac{1}{6} \sum_{p=1}^8 w_k^p \, \mathrm{I}^3_p[Q] \right) \right]\hspace{-1mm},&&
\eea
where the monomials $\mathrm{I}^1[Q]$, $\{ \mathrm{I}^2_p[Q]\}$, $\{ \mathrm{I}^3_p[Q] \}$ are listed in Table \ref{tab1}. In Eq. (\ref{eq4}), $\mathcal{Z}_k$ is given by a combination of twelve terms: given that these are all the monomials of degree three or less that are consistent with the symmetries of the  model (\ref{ghea}), Eq. (\ref{eq4}) constitutes the most general form of the third-order expansion of $\mathcal{Z}_k$  \cite{temesvari2008almeida}. In the limit of zero magnetic field additional symmetries appear, and the set of admissible monomials is reduced \cite{kotliar1983one} to $\mathrm{I}^2_1[Q]$ and $\mathrm{I}^3_1[Q]$. \\

Plugging Eq. (\ref{eq4}) into the RG equation (\ref{eq3}), we obtain
\bw
\bea\label{eq6}
\mathcal{Z}_{k+1}[Q]  & = & \exp\left( \frac{\beta^2}{4} \text{Tr}[Q^2]  - \frac{2 \, s_k }{C^{1/2}} \mathrm{I}^1[Q] - \frac{1}{2} \sum_{p=1}^3 \frac{2 \, r^p_k}{C}  \mathrm{I}^2_p[Q] - \frac{1}{6} \sum_{p=1}^8 \frac{2 \, w^p_k}{C^{3/2}} \mathrm{I}^3_p[Q] \right) \times \\ \nn
&& \times \int [dP]\exp\Bigg( - \frac{1}{2} \sum_{a<b,c<d} P_{ab} \mathscr{M}_{ab,cd}[Q] P_{cd}\Bigg),
\eea
where $\mathscr{M}_{ab,cd}$ is a $n(n-1)/2 \times n(n-1)/2$ matrix given by
\be\label{eq5}
\mathscr{M}_{ab,cd}[Q]  =  \mathscr{F}_{ab,cd} + \mathscr{G}_{ab,cd}[Q],
\ee
and
\bea\label{eq8}
\mathscr{F}_{ab,cd} & = &   \frac{4 \, r^1_k}{C} \left( \de_{ac} \de_{bd} + \de_{bc} \de_{ad} \right) + \frac{2 \, r^2_k}{C}\left(\de_{ac} + \de_{bc} + \de_{ad} + \de_{bd}\right) + \frac{8 \, r^3_k}{C},\\  \label{eq7}
\mathscr{G}_{ab,cd}[Q] & = & \frac{1}{6} \Bigg\{ \frac{12 \, w^1_k}{C^{3/2}} \left( \de_{ad} Q_{bc} +  \de_{bd} Q_{ac} +  \de_{ac} Q_{bd} +  \de_{bc} Q_{ad}\right)  + \frac{24 \, w^2_k}{C^{3/2}} \left( \de_{ac} \de_{bd} + \de_{bc} \de_{ad}   \right) Q_{cd}  + \\ \nn
&& + \frac{4 \, w^3_k}{C^{3/2}} \Bigg[  \left( \de_{ac} \de_{bd}  + \de_{bc} \de_{ad} \right) \sum_{e=1}^n\left( Q_{ce} + Q_{de} \right) +   \left( \de_{ac} +\de_{bc} +\de_{ad} +\de_{bd} \right) Q_{cd} + \\ \nn
&& + \de_{ac} Q_{bc} +\de_{bc} Q_{ac}+\de_{ad} Q_{bd} +\de_{bd} Q_{ad} \Bigg]  + \frac{8 \, w^4_k}{C^{3/2}}\left[ \left( \de_{ac} \de_{bd} + \de_{bc} \de_{ad} \right) \mathrm{I}^1[Q] + 2 \left( Q_{ab} + Q_{cd} \right) \right] + \\ \nn
&& + \frac{4 \, w^5_k}{C^{3/2}} \Bigg[ \sum_{e=1}^n \left( \de_{ac} Q_{de} + \de_{bc} Q_{de}+  \de_{ad} Q_{ce} +  \de_{bd} Q_{ce} + \de_{ac} Q_{be} + \de_{bc} Q_{ae}  + \de_{ad} Q_{be} + \de_{bd} Q_{ae} \right) + \\ \nn
&& + Q_{ac} + Q_{bc} + Q_{ad} + Q_{bd}  \Bigg] + \frac{12 \, w^6_k}{C^{3/2}} \sum_{e=1}^n \left( \de_{ac} Q_{ce} +\de_{bc} Q_{ce}+\de_{ad} Q_{de} +\de_{bd} Q_{de} \right) + \\ \nn
&& + \frac{4 \, w^7_k}{C^{3/2}} \left[ \left( \de_{ac} + \de_{bc} + \de_{ad} + \de_{bd} \right) \mathrm{I}^1[Q] + 2 \sum_{e=1}^n \left( Q_{ae} + Q_{be}  + Q_{ce} + Q_{de} \right) \right] + \frac{48 \, w^8_k}{C^{3/2}} \mathrm{I}^1[Q]\Bigg\}.
\eea
\ew

Note that in Eq. (\ref{eq5}) we split the matrix $\mathscr{M}$ into a $Q$-independent term $\mathscr{F}$ and a $Q$-dependent term $\mathscr{G}$, the latter  being proportional to the coefficients $\{ w^p_k \}$: indeed, in what follows we will make the assumption that the coefficients $\{ w^p_k \}$ in Eq. (\ref{eq4}) can be treated in perturbation theory, and the form (\ref{eq5}) will be convenient for solving the RG equation (\ref{eq6}) with an expansion in powers of $\{ w^p_k \}$. To set up this expansion, let us first compute the inverse of $\mathscr{F}$. This can be sought as a combination of the terms that appear in Eq. (\ref{eq8}):
\bea\nn
\mathscr{F}^{-1}_{ab,cd} &=&\\ \label{eq15}
 c_1 \left( \de_{ac} \de_{bd} + \de_{bc} \de_{ad} \right) + c_2 \left(\de_{ac} + \de_{bc} + \de_{ad} + \de_{bd}\right) +  c_3,&& 
\eea
and the coefficients $c_1, c_2, c_3$ can be determined by solving the equation $\sum_{c<d} \mathscr{F}_{ab,cd} \mathscr{F}^{-1}_{cd,ef} = \de_{ae} \de_{bf}$. The result is 
\bea \label{eq22}
c_1&  = & \frac{C}{4 r^1_k}, \\\label{eq23}
c_2 & = & - \frac{C r^2_k}{4 r^1_k\left(2 r^1_k + (n-2) r^2_k \right) },\\\label{eq24}
c_3 & = & \frac{ C( (r^2_k )^2 \hspace{-1mm}-\hspace{-1mm} 2\,  r^1_k r^3_k + n \, r^2_k r^3_k )}{2\, r^1_k \left(2 \, r^1_k \hspace{-1mm}+\hspace{-1mm} (n \hspace{-1mm}-\hspace{-1mm} 2) r^2_k \right) \left( r^1_k \hspace{-1mm}+\hspace{-1mm} (n-1) \left( r^2_k \hspace{-1mm}+\hspace{-1mm} n \, r^3_k \right)\right)}.
\eea
The matrix $\mathscr{F}^{-1}$ can now be used to compute the second factor in Eq. (\ref{eq6}):
\bea\label{eq9}\nn
\int [dP]\exp\Bigg( - \frac{1}{2} \sum_{a<b,c<d} P_{ab} \mathscr{M}_{ab,cd}[Q] P_{cd}\Bigg) & = & \\ \nn
\frac{\alpha}{\sqrt{\det(\mathscr{M}[Q])}} & = & \\ \nn
\alpha \exp\left( -\frac{1}{2} \mathrm{Tr} \log\left(1 + \mathscr{F}^{-1} \cdot \mathscr{G}[Q] \right) \right) & = & \\ \nn
\alpha \exp\Bigg(- \frac{1}{2} \mathrm{Tr}\Bigg[ \mathscr{F}^{-1} \cdot \mathscr{G}[Q]  -\frac{1}{2} \left(\mathscr{F}^{-1} \cdot \mathscr{G}[Q]\right)^2 + &&\\ 
 + \frac{1}{3} \left(\mathscr{F}^{-1} \cdot \mathscr{G}[Q]\right)^3 + O(Q^4) \Bigg] \Bigg),&&
\eea
where in Eq. (\ref{eq9}) $\alpha$ denotes a $Q$-independent factor, and in the third line the symbol `$1$' in the logarithm denotes the $n(n-1)/2 \times n(n-1)/2$ identity matrix.

Given the complex form (\ref{eq7})  of the matrix $\mathscr{G}$, the calculation of the matrix products and traces in Eq. (\ref{eq9}) is  involved: for example, if we denote by a `term' in Eqs. (\ref{eq8}), (\ref{eq7}) the sum of multiple addends related to each other by the permutation $a \leftrightarrow b$, $c \leftrightarrow d$---such as $\de_{ac} + \de_{bc} + \de_{ad} + \de_{bd}$---then a direct evaluation of $\mathrm{Tr}[(\mathscr{F}^{-1}\mathscr{G})^3]$ involves the computation of $ (3 \times 14)^3 = 74,088$ matrix traces. To handle this computation, we developed a symbolic manipulation program \cite{note3}  that enabled us to compute explicitly the right-hand side (RHS) of Eq. (\ref{eq9}): Sums over replica indexes are computed in a symbolic way, and the result depends explicitly on the number of replicas $n$, allowing for a straightforward analytic continuation \cite{mezard1987spin} for $n\rightarrow0$. Importantly, the result of this calculation shows that the RHS of Eq. (\ref{eq9}) is a linear combination of the monomials listed in Table \ref{tab1}. This implies that $\mathcal{Z}_{k+1}[Q]$ and $\mathcal{Z}_k[Q]$ are both given by a linear combination of the same monomials, hence the RG equation (\ref{eq3}) can be solved consistently in perturbation theory. In particular, Eqs.  (\ref{eq4}), (\ref{eq6}), (\ref{eq9}) imply a set of recursive equations which relate $s_{k+1}$,  $\{r_{k+1}^p\}$, $\{ w_{k+1}^p \}$ to $s_k$, $\{r_k^p\}$, $\{w_k^p \}$: the $n \rightarrow 0$ limit of these equations is given by Eqs. 
%(\ref{eq1.1})-(\ref{eq3.8})
(S1)-(S12) in Section 
%\ref{secs1}
S1 of the Supplemental Material.
In addition, in Section 
%\ref{secs2}
S2 we discuss how to write the RG coefficients $s_k$, $\{r_k^p\}$, $\{w_k^p \}$ in terms of the interaction-decay exponent, the temperature and the magnetic-field strength  to study, for example, how the RG flow or the FP domains of attraction are related to the physical parameters of the model. 

\subsection{Structure of the critical fixed point}\label{sec1.2}

We will now study the existence of a  FP associated with a spin-glass transition and determine its general features. To do so, we consider the spin-glass susceptibility \cite{katzgraber2009study,leuzzi2009ising}
\be\label{eq10}
\chi_{\rm SG} \equiv \frac{1}{2^k} \sum_{i,j=1}^{2^k} \mathbb{E}\big[\left(\langle S_i S_j \rangle - \langle S_i \rangle \langle S_j \rangle\right)^2\big],
\ee 
where $\langle \, \rangle$ denotes the Boltzmann average with Hamiltonian (\ref{ghea}). The susceptibility (\ref{eq10}) can be rewritten as \cite{green1983upper}
\be\label{eq11}
\chi_{\rm SG} = 2^k \, \mathbb{E}[\langle q_{12}^2 \rangle - 2 \langle q_{12} q_{13} \rangle + \langle q_{12} q_{34} \rangle ],
\ee 
where in what follows the average $\langle \,  \rangle$ of a function of multiple replicas denotes the  average over all replicas  \cite{mezard1987spin}, each replica having an independent Boltzmann measure with Hamiltonian (\ref{ghea}).  
By using Eq. 
%(\ref{eqs11})
(S20), the overlap averages in the RHS of Eq. (\ref{eq11}) can be rewritten as follows:
\bea\label{eq12}
\mathbb{E}[\langle q_{12}^2 \rangle] & = & C^{-k} \lim_{n \rightarrow 0} \overline{Q_{12}^2},\label{eq16}\\
\mathbb{E}[\langle q_{12}q_{13}\rangle] & = & C^{-k} \lim_{n \rightarrow 0} \overline{Q_{12}Q_{13}},\\  \label{eq17}
\mathbb{E}[\langle q_{12}q_{34}\rangle] &  = & C^{-k} \lim_{n \rightarrow 0}  \overline{Q_{12}Q_{34}},
\eea
where $\overline{\phantom{A}}$ denotes the average with respect to $\mathcal{Z}_k[Q]$. Neglecting $O(w)$ terms and using Eqs. (\ref{eq4}), (\ref{eq8}), we obtain 
\bea\nn\label{eqz}
\mathcal{Z}_k[Q] & = & \\ 
\exp\left[-\left( s_k \, \mathrm{I}^1[Q] + \frac{C}{4} \sum_{a<b,c<d} Q_{ab} \mathscr{F}_{ab,cd} Q_{cd} \right) \right]. &&
\eea
By using Eq. (\ref{eqz}) and standard Gaussian-integration rules \cite{justin2002quantum}, we have 
\bea \label{eq18}
\overline{Q_{12}^2} & = & \left( \frac{4 s_k}{C}  \right)^2 \mathscr{H}^2 + \frac{2}{C} \mathscr{F}^{-1}_{12,12}\\ \nn
& = & \left( \frac{4 s_k}{C}  \right)^2 \mathscr{H}^2+ \frac{2}{C}(c_1 + 2 c_2 + c_3),\\  \label{eq19}
\overline{Q_{12}Q_{13}} & = & \left( \frac{4 s_k}{C}  \right)^2 \mathscr{H}^2 + \frac{2}{C} \mathscr{F}^{-1}_{12,13} \\ \nn
&  = &  \left( \frac{4 s_k}{C}  \right)^2 \mathscr{H}^2 + \frac{2}{C} (c_2 + c_3), \\ \label{eq20}
\overline{Q_{12}Q_{34}} & = & \left( \frac{4 s_k}{C}  \right)^2 \mathscr{H}^2 + \frac{2}{C} \mathscr{F}^{-1}_{12,34} \\ \nn
&=& \left( \frac{4 s_k}{C}  \right)^2 \mathscr{H}^2 + \frac{2}{C} c_3, 
\eea
where in Eqs. (\ref{eq18})-(\ref{eq20}) we set $\mathscr{H} \equiv \sum_{a<b} \mathscr{F}^{-1}_{12,ab}$ and we used Eq. (\ref{eq15}). 
Putting together Eqs. (\ref{eq21}), (\ref{eq22}),  (\ref{eq11}), (\ref{eq12})-(\ref{eq17}), (\ref{eq18})-(\ref{eq20}), we obtain the expression for the spin-glass susceptibility as a function of $r^1_k$:
\be\label{eq25}
\chi_{\rm SG} = \left(\frac{2}{C} \right)^{k+1} \lim_{n \rightarrow 0} c_1 = \left(\frac{2}{C} \right)^{k+1} \lim_{n \rightarrow 0}  \frac{C }{4 r^1_k}.
\ee
Since we are assuming that $\chi_{\rm SG}$ diverges at the critical point, if any, the coefficient $r^1_k$ must have the following behavior
\be\label{eq90}
r^1_k \rightarrow r^1_\ast \text{ for } k \rightarrow \infty,  T = T_c \text{ with } r^1_\ast \text{ finite.}
\ee
Indeed, if Eq. (\ref{eq90}) did not hold, then according to Eq. 
%(\ref{eq2.1})
(S2) $r^1_k$ would diverge like $r^1_k \sim (2/C)^k$, and Eq. (\ref{eq25}) would imply that $\chi_{\rm SG}$ is finite. 
We recall that the conclusion above on the large-$k$ behavior of $r^1_k$ at the critical point has been derived by neglecting the $O(w)$ terms in Eq. (\ref{eq4}): given that our RG analysis is based on the working hypothesis that perturbation theory is well-behaved, retaining the $O(w)$ contributions would result in a perturbative correction to $r^1_\ast$, without changing the qualitative behavior (\ref{eq90}).

We have thus derived a first property of the critical FP: at the critical temperature $T = T_c$, we have  $r^1_k \rightarrow r^1_\ast$ for $k \rightarrow \infty$, with $r^1_\ast$ finite. Importantly, Eq. 
%(\ref{eq2.1})
(S2) shows that if $T \neq T_c$, then $r^1_k$ diverges like $r^1_k \sim (2/C)^k$ for large $k$: setting $T = T_c$, the projection of the RG flow along the direction associated with $r^1_k$ is set to zero, and the divergence of $r^1_k$ is removed \cite{wilson1974renormalization}. 

The critical FP can now be completely characterized as follows. According to Eqs. 
%(\ref{eq1.1}), (\ref{eq2.2}), (\ref{eq2.3})
(S1), (S3), (S4), the coefficients $s_k, r^2_k, r^3_k$ diverge for large $k$: we cannot remove these divergences like we did for the coefficient $r^1_k$, because the only free parameter in the model---the temperature---has already been fixed  to eliminate the divergence of $r^1_k$. Thus, the only possible FP for these coefficients is $s_k= s_\ast = \pm \infty$, $ r^2_k = r^2_\ast = \pm \infty$, $r^3_k = r^3_\ast =  \pm \infty$. Finally, given our perturbative working hypothesis,  the coefficients $\{ w^p_k \}$ are assumed to be small, thus they must converge to a finite FP $w^p_k = w^p_\ast$, $p=1, \ldots,8$.

Given the critical-FP structure above, the critical values $\{ c_p^\ast \}$ of the coefficients $\{ c_p \}$ are finite despite the fact that $s_\ast$, $r^2_\ast$ and $r^3_\ast$ are infinite: 
\bea\label{eq29}
 c_1^\ast& =&  \frac{C}{4 r^1_\ast},\\ \label{eq30}
c_2^\ast& = &\frac{C}{8 r^1_\ast},\\\label{eq31}
c_3^\ast &= &\frac{C}{4 r^1_\ast},
\eea
where to obtain Eqs. (\ref{eq29})-(\ref{eq31}) we used Eqs. (\ref{eq22})-(\ref{eq24}) and the conditions $r^2_\ast = \pm \infty$,  $r^3_\ast = \pm \infty$, $n \rightarrow 0$. Importantly, the finiteness of $\{ c_p^\ast \}$ ensures that the FP equations for $r^1_\ast$, $\{ w^p_\ast\}$ are well posed because they involve only finite terms---see Eqs. 
%(\ref{eq2.1}), (\ref{eq3.1})-(\ref{eq3.8})
(S2), (S5)-(S12). 

\subsection{Solution of the fixed-point equations }\label{sec1.3}

We will now determine explicitly the critical FP.  First of all, the RG Eqs. 
%(\ref{eq2.1}), (\ref{eq3.1})-(\ref{eq3.8})
(S2), (S5)-(S12) possess a trivial FP
\be\label{eq47}
r^1_\ast = \frac{\beta^2}{2(2/C-1)}, \; w^1_\ast = \cdots = w^8_\ast = 0. 
\ee
As shown in Appendix \ref{app3.1}, in the mean-field region $1/2 < \sigma \leq 2/3$ this FP is stable, and it is thus associated with  the existence of a physical spin-glass transition in a magnetic field. 

In non-mean-field region $2/3 <\sigma < 1$ the FP (\ref{eq47}) is unstable, see Appendix \ref{app3.1}, thus other FPs must be considered. To this end, we observe that the FP equations can be simplified if $\sigma$ lies in the neighborhood of the threshold value $\sigma = 2/3$. To show this, we set
\be\label{eq26}
\epsilon \equiv \sigma  - 2/3,
\ee 
and we observe \cite{castellana2011renormalization} that for small $\epsilon$ the RG Eqs. 
%(\ref{eq2.1}), (\ref{eq3.1})-(\ref{eq3.8})
(S2), (S5)-(S12) are of the form $w^p_\ast \epsilon = O((w^p_\ast)^3)$, implying that $(w^p_\ast)^2 = O(\epsilon)$. Hence, we set 
\be\label{eq27}
w^p_\ast = \beta^3 \left[ \frac{\log 2}{(2^{1/3}-1)^3}\right] ^{1/2} \, \omega_p \sqrt{\epsilon},
\ee
where  $\{ \omega_p \}$ are  coefficients of order unity \bibnote[note2]{We have chosen to incorporate the factor  $\beta^3 (\log 2 /(2^{1/3}-1)^3)^{1/2}$ in the definition (\ref{eq27}) because this factor ensures that the FP equations for $\{ \omega_p \}$, Eqs. (\ref{eqf3.1})-(\ref{eqf3.8}), have integer coefficients independent of $\beta$, thus allowing for an exact computation of their Gr\"{o}bner basis.}. Also,  Eq. (\ref{eq27}) and the recursive equation 
%(\ref{eq2.1})
(S2) imply that the FP for $r^1_\ast$ reads
\be\label{eq28}
r^1_\ast = \frac{\beta^2}{2(2^{1/3}-1)} + \rho \, \epsilon,
\ee
where $\rho$ is a coefficient of order unity. Note that the mean-field FP (\ref{eq47}) can be obtained from Eqs. (\ref{eq27}), (\ref{eq28}) by setting $\epsilon = 0$: hence, the FP (\ref{eq27}), (\ref{eq28}) can be regarded as a perturbation of the mean-field FP.  We will now determine the critical FP, if any. To this end, we  use Eqs. (\ref{eq27}), (\ref{eq28}), (\ref{eq29})-(\ref{eq31}), and we obtain \bibnote[note3]{The calculation is detailed in the online Mathematica \cite{wolfram2014mathematica} notebook \texttt{symbolic\char`_computation.nb}, which is available as an ancillary file.} that the FP condition for Eqs. 
%(\ref{eq2.1}), (\ref{eq3.1})-(\ref{eq3.8})
(S2), (S5)-(S12) implies the following set of polynomial equations for $\rho$, $\{ \omega_p \}$
\bw
\bea\label{eqf2.1}
4 (2^{1/3}-1)^2 \rho + 2^{1/3} \beta^2 \log 2 \left(-4 \omega_1^2+16 \omega_1 \omega_2-11 \omega_2^2+4\right) & = & 0, \\ 
\label{eqf3.1}
14 \omega_1^3-36 \omega_1^2 {\omega_2}+18 {\omega_1} \omega_2^2+\omega_2^3 + 6 {\omega_1} & = &0,\\
\label{eqf3.2}
{\omega_2} \left( 3 + 12 \omega_1^2 -30 {\omega_1} \omega_2+17 \omega_2^2 \right) & = & 0,\\
\label{eqf3.3}
18 \omega_1^3-4 \omega_1^2 (15 {\omega_2}+2 {\omega_3})+8 {\omega_1} {\omega_2} (9 {\omega_2}+4 {\omega_3})-11 \omega_2^2 (3 {\omega_2}+2 {\omega_3})- 6{\omega_3} & = & 0,\\
\label{eqf3.4}
6 \omega_1^3+8 \omega_1^2 ({\omega_3}+2 {\omega_4})-{\omega_1} {\omega_2} (15 {\omega_2}+12 {\omega_3}+64 {\omega_4})+\omega_2^2 (12 {\omega_2}+7 {\omega_3}+44 {\omega_4}) + 12 {\omega_4}& = & 0, \\
\label{eqf3.5}
18 \omega_1^3-3 \omega_1^2 (27 {\omega_2}+4 {\omega_3})+6 {\omega_1} \left(6 \omega_2^2-3 {\omega_2} {\omega_3}-\omega_3^2\right)+{\omega_2} \big(18 \omega_2^2+24 {\omega_2} {\omega_3} +7 \omega_3^2\big) +  18 {\omega_5}  & = & 0,\\
\label{eqf3.6}
72 \omega_1^3-243 \omega_1^2 {\omega_2}+18 {\omega_1} \left(12 \omega_2^2+{\omega_2} {\omega_3}+\omega_3^2\right)-63 \omega_2^3-18 \omega_2^2 {\omega_3}-15 {\omega_2} \omega_3^2+4 \omega_3^3- 54 {\omega_6}& = & 0,\\
\label{eqf3.7}
9 \omega_1^3+3 \omega_1^2 (6 {\omega_2}+7 {\omega_3}+8 {\omega_4})+3 {\omega_1} \left[3 \omega_2^2+4 {\omega_2} ({\omega_3}+4 {\omega_4})+{\omega_3} (3 {\omega_3}+16 {\omega_4})\right]-36 \omega_2^3 -3 \omega_2^2 (11 {\omega_3}+& \\ \nn 
+20 {\omega_4})-2 {\omega_2} {\omega_3} (5 {\omega_3}+24 {\omega_4})-\omega_3^2 ({\omega_3}-8 {\omega_4})- 36 {\omega_7} & = & 0,\\
\label{eqf3.8}
63 \omega_1^3-9 \omega_1^2 [3 {\omega_2}-5 ({\omega_3}+4 {\omega_4})]-9 {\omega_1} ({\omega_3}+4 {\omega_4}) (2 {\omega_2}-{\omega_3}-4 {\omega_4})-36 \omega_2^3-36 \omega_2^2 ({\omega_3}+2 {\omega_4}) -3 {\omega_2} \times &\\ \nn
\times \left(5 \omega_3^2+24 {\omega_3} {\omega_4}+48 \omega_4^2\right)-{\omega_3} \left(\omega_3^2+12 {\omega_3} {\omega_4}-48 \omega_4^2\right) - 216 {\omega_8} & = & 0.
\eea
\ew

The solution to the equations above can be determined by solving Eqs. (\ref{eqf3.1})-(\ref{eqf3.8}) for $\{ \omega_p \}$ first, and then substituting the solution into Eq. (\ref{eqf2.1}) to obtain $\rho$. Being a system of polynomial equations of the third degree, in general Eqs. (\ref{eqf3.1})-(\ref{eqf3.8}) have multiple solutions. To find all solutions, we determined the Gr\"{o}bner basis (GB) of the system of polynomials (\ref{eqf3.1})-(\ref{eqf3.8}): the GB is given by a set of polynomials in  $\{ \omega_p \}$, and the roots of Eqs. (\ref{eqf3.1})-(\ref{eqf3.8}) coincide with those of the GB \cite{froberg1997introduction}. 
Given that Eqs. (\ref{eqf3.1})-(\ref{eqf3.8}) have integer coefficients, their GB can be computed exactly, and it reads
\bw
\bea\label{eq80}
\omega_8 \left(28 \omega_8^2+3\right) \left\{ 16 \left[ 80 \left(8656 \omega_8^2+933\right) \omega_8^2+2367\right] \omega_8^2+3267\right \},\\
\omega_7-3 \omega_8\label{eq81}, \\ 
1877310171 \omega_6+32100226170880 \omega_8^7+2352496281600 \omega_8^5+641548992 \omega_8^3+8790294330 \omega_8, \\
625770057 \omega_5-8 \omega_8 \left(28 \omega_8^2+3\right) \left(71652290560 \omega_8^4-2425923360 \omega_8^2+261352389\right), \\
625770057 \omega_4+8025056542720 \omega_8^7+588124070400 \omega_8^5+160387248 \omega_8^3+1258918497 \omega_8, \\
625770057 \omega_3+2 \omega_8 \left(8025056542720 \omega_8^6+588124070400 \omega_8^4+160387248 \omega_8^2+1258918497\right), \\
1877310171 \omega_2-16 \omega_8 \left(28 \omega_8^2+3\right) \left(71652290560 \omega_8^4-2425923360 \omega_8^2+261352389\right), \\\label{eq82}
625770057 \omega_1-2 \omega_8 \left(8025056542720 \omega_8^6+588124070400 \omega_8^4+160387248 \omega_8^2+2510458611\right),
\eea
\ew
where every line in Eqs. (\ref{eq80})-(\ref{eq82}) corresponds to an element of the GB. The first GB element, Eq.  (\ref{eq80}), depends only on $\omega_8$, and its complete set of roots can be determined exactly. Since the GBs (\ref{eq81})-(\ref{eq82}) are linear in $\omega_1, \ldots, \omega_7$, to every solution for $\omega_8$ corresponds a unique value for $\omega_1, \ldots, \omega_7$. As a consequence, the GB method allows for extracting the full set of exact solutions. The numerical root values for $\omega_8$ read $\omega_8 = 0, \,  0.159676 \pm 0.167743 \, \imi, -0.159676 \pm 0.167743 \, \imi, \, \pm 0.327327 \, \imi, \, \pm 0.320162 \, \imi$. First, it is straightforward to show that the trivial root $\omega_8 = 0$ corresponds to the FP
\be\label{eq33}
\rho = - \beta^2 \frac{2^{1/3} \log 2}{ (2^{1/3}-1)^2}, \;\; \omega_1 = \cdots = \omega_8  = 0,
\ee
and that this FP is unstable, thus it does not correspond to a spin-glass transition---see Appendix \ref{app3.2} for details. Second, all other roots for $\omega_8$ have a nonzero imaginary part. This implies that, since the initial condition $\mathcal{Z}_0[Q]$ of the RG transformation (\ref{eq3}) is real and since the RG transformation maintains reality, none of the FPs in the non-mean-field region $\sigma = 2/3 + \epsilon $ are physically accessible by the RG flow.

\section{Conclusions}\label{sec2}

Establishing the existence of a phase transition in non-mean-field spin glasses with an external magnetic field is an open problem which has been attracting growing interest in recent years \cite{petit1999ordering,jonsson2005dynamical,jorg2008behavior,young2004absence,banos2012thermodynamic,katzgraber2009study,leuzzi2009ising}. Indeed, the occurrence of such a transition is believed to be related to the structure of the low-temperature phase of spin glasses---a central topic in statistical physics of disordered systems \cite{fisher1986ordered}. 
Among non-mean-field models of spin glasses, the hierarchical Edwards-Anderson model \cite{franz2009overlap} is the simplest non-mean-field spin-glass system where the hierarchical structure \cite{dyson1969existence} of spin interactions allows for a natural implementation of renormalization-group (RG) techniques:
recent studies showed that these RG methods provide a novel way of understanding the thermodynamical properties of the model \cite{castellana2011renormalization,castellana2011real,angelini2013ensemble,angelini2014spin}. 

In this paper, we studied the existence of a phase transition for the hierarchical Edwards-Anderson model with an external magnetic field (HEAM): We used a RG approach based on the replica method \cite{castellana2011renormalization}, and we analyzed in perturbation theory the RG flow of the average replicated partition function with fixed overlap. In this approach, the replicated partition function is described by a set of twelve  parameters: given the complex algebraic structure of the resulting RG theory, we developed a novel symbolic computation method which allowed us for extracting the RG equations to lowest order in perturbation theory. 

In the absence of a magnetic field, the HEAM possesses a mean-field region $1/2 < \sigma \leq 2/3$ where order-parameter fluctuations vanish, and a non-mean-field region $2/3 < \sigma <1$ characterized by nonzero order-parameter fluctuations, where $\sigma$ is a parameter tuning the interaction decay with distance \cite{castellana2011renormalization}. We investigated the occurrence of a spin-glass transition in the HEAM by studying the existence of a stable RG fixed point (FP) in both these parameter regimes. 
In the mean-field region we found a stable critical FP which corresponds to a physical spin-glass transition. In the non-mean-field region, we investigated the FPs contiguous to the mean-field one with a perturbative expansion in $\epsilon = \sigma - 2/3$. The resulting FP equations are a system of polynomial equations 
of the third degree which possess, in principle, multiple FP solutions. 
In this regard, two previous RG studies investigated the perturbative FPs in the non-mean-field region of short-range spin-glass models in a magnetic field. First, a set of unstable solutions of the FP equations was found \cite{pimentel2002spin}, but these solutions  were not shown to be the complete set of roots of the FP equations \cite{pimentel2014private}. 
Second, the complete set of FPs has been extracted in a reduced framework where only a subset of the order-parameter modes is retained \cite{bray1980renormalisation,moore2011disappearance}. By using a Gr\"{o}bner-basis method for systems of polynomial equations \cite{froberg1997introduction}, here we computed exactly all perturbative FPs for the full set of RG equations in the non-mean-field region of the HEAM. Our analysis shows that that all potentially stable FPs have a nonzero imaginary part. Given that in the RG transformation the initial values of the coefficients are real, and given that the RG transformation maintains reality, our results constitute, to the best of our knowledge, the first demonstration for a spin glass in a field that the there exists no perturbative FP in the non-mean-field region.

To interpret the absence of a perturbative FP in the non-mean-field region, several scenarios can be considered \cite{bray1980renormalisation}. 
First, the absence of a perturbative FP may imply that there is no spin-glass transition in the non-mean-field region \cite{moore2011disappearance,katzgraber2009study}. 
In this regard, we recall that the addition of a random magnetic field in the ferromagnetic Ising model increases the lower critical dimension from $d = 1$ in the zero-field case to $d= 2$ in a finite field \cite{imbrie1984lower}: along these lines, the inclusion of a random field in the hierarchical Edwards-Anderson model may decrease the value of $\sigma$ corresponding \cite{katzgraber2009study} to the lower critical dimension from $\sigma = 1$ in zero field to $\sigma = 2/3$ in a finite field, thus providing a possible explanation for the absence of a transition for $2/3<\sigma < 1$. 
A second possibility is that a  FP in a field for $\sigma = 2/3 + \epsilon$ exists, but this FP may not be contiguous to the mean-field one, thus it may not be found with the perturbative approach used here: this hypothesis is in line with the fact that perturbative approaches are generally not well understood in spin glasses \cite{castellana2014renormalization,castellana2011renormalization,castellana2011real}. A third possibility is that there is a phase transition in a field, but this transition is not associated with a FP within the replica RG method. 
As a future direction, the last two possibilities may be investigated with recent real-space RG approaches developed for the hierarchical Edwards-Anderson model \cite{castellana2011real,angelini2013ensemble} which do not rely on perturbation theory, nor they make use of the replica formalism.  
Non-perturbative effects could be also studied with Monte Carlo (MC) simulations. 
In this regard, recent MC studies focused on four-dimensional short-range spin glasses \cite{banos2012thermodynamic} and one-dimensional long-range spin glasses  which correspond to short-range models with $d = 4$ \cite{leuzzi2009ising}. These works hinted at the existence of a transition in a magnetic field in the non-mean-field region, a picture which is at variance with the perturbative results provided here for the HEAM. 
Non-perturbative effects could be pinned down directly in the HEAM by means of a systematic comparison between the perturbative RG predictions provided here and MC simulations \cite{castellana2015non}: indeed, the RG flow  of the replicated partition function $Z_k \rightarrow Z_{k+1}$ could be directly investigated numerically  by  computing  the moments of the overlap distribution for different system sizes $2^k$.
If the numerics provided evidence for a transition in the non-mean-field region, one could then probe directly the nature of the FP corresponding to such transition by characterizing numerically its critical exponent $\nu$  \cite{wilson1974renormalization}. If the MC estimate of $\nu$ was found to be close to the classical value $\nu_{\rm{cl}}  = 1/(2 \sigma-1)$ \cite{castellana2011renormalization} for $\sigma \gtrsim 2/3$, then the spin-glass transition resulting from the numerics could be associated with a FP which is a perturbation of the mean-field one. Conversely, a strong discrepancy  between the MC estimate of $\nu$ and $\nu_{\rm{cl}}$ for $\sigma \gtrsim 2/3$ would hint at the existence of a non-perturbative FP lying outside the domain of attraction of the mean-field FP.
Finally, a further natural way to examine non-perturbative effects consists in studying the behavior of the $\epsilon$-expansion to large orders \cite{castellana2011renormalization}: to this end, the symbolic computation method introduced in this paper may be helpful in setting up a fully automated, large-order $\epsilon$-expansion, which may be useful for understanding the limits of perturbation theory in spin glasses.

\acknowledgments{
M. C. is grateful to G. Parisi, S. Franz and M. A. Moore for useful comments and discussions.
C. B. is grateful to T. Rizzo for useful insights into the automated symbolic computation and to D. Lichtblau at Wolfram Research for discussions on the solution of the fixed-point equations. 
Research supported in part by the European Research Council through grant agreement no. 247328--CriPherasy project, by NSF Grants PHY--0957573, PHY--1305525 and CCF--0939370, by the Human Frontiers Science Program, by the Swartz Foundation, and by the W. M. Keck Foundation.
The symbolic calculations presented in this article were performed on computational resources supported by the Lewis-Sigler Institute for Integrative Genomics at Princeton University. 
}

\appendix

\section{Stability of the trivial fixed point (\ref{eq47}) in the mean-field region}\label{app3.1}

To study the stability of the FP (\ref{eq47}), we set $\vec{x}_k = (x_k^1, \ldots, x_k^9) \equiv (r_k, w^1_k, \cdots, w^8_k)$ and we linearize the transformation $\vec{x}_k \rightarrow \vec{x}_{k+1}$ implied by Eqs. 
%(\ref{eq2.1}), (\ref{eq3.1})-(\ref{eq3.8})
(S2), (S5)-(S12) in the neighborhood of the FP $x_\ast = (r^1_\ast, 0, \ldots, 0)$: the FP is stable if the matrix 
\be\label{eq49}
\mathcal{M}_{ij} \equiv \left. \frac{\partial x_{k+1}^i}{\partial x_k^j}\right|_{\vec{x}_\ast},
\ee
has not more than one eigenvalue larger than one \cite{wilson1974renormalization}. By using Eqs. 
%(\ref{eq2.1}), (\ref{eq3.1})-(\ref{eq3.8})
(S2), (S5)-(S12), (\ref{eq49}), it is straightforward to obtain the eigenvalues of $\mathcal{M}$, which read $\lambda_1 = 2/C, \lambda_2 = \cdots = \lambda_9 = 2/C^{3/2}$. It follows that in the mean-field region $1/2 < \sigma \leq 2/3$ only $\lambda_1$ is larger than one, and the trivial FP (\ref{eq47}) is stable. In the non-mean-field region $\lambda_2, \cdots, \lambda_9$ are all larger than one, and the FP (\ref{eq47}) is unstable. \\

\section{Instability of the trivial fixed point (\ref{eq33})}\label{app3.2}

We will show that the FP (\ref{eq33}) is unstable by proceeding along the lines of Appendix \ref{app3.1}: since we want to study the RG flow in the neighborhood of a FP of the form (\ref{eq27}), (\ref{eq28}), we set 
\bea\label{eq34}
r^1_k & = &\frac{\beta^2}{2(2^{1/3}-1)} + \rho_k \, \epsilon,\\\label{eq36}
w^p_k & = & \beta^3 \left[\frac{\log 2}{(2^{1/3}-1)^3}\right]^{1/2} \, \omega_{p \, k} \sqrt{\epsilon}. 
\eea
We then  introduce the vector $\vec{y}_k  = (y_k^1, \ldots, y_k^9) \equiv (\rho_k, \omega_{1\, k}, \ldots, \omega_{8 \, k})$ and the RG transformation  $\vec{y}_k \rightarrow \vec{y}_{k+1}$ implied by Eqs. 
%(\ref{eq2.1}), (\ref{eq3.1})-(\ref{eq3.8})
(S2), (S5)-(S12), (\ref{eq34}), (\ref{eq36}). Then, we consider the FP $\vec{y}_\ast = (\rho, \omega_1, \cdots, \omega_8)$, where $\rho, \omega_1, \cdots, \omega_8$ are given by Eq. (\ref{eq33}): this FP is stable if the matrix
\be\label{eq35}
\mathcal{N}_{ij} \equiv \left. \frac{\partial y_{k+1}^i}{\partial y_k^j}\right|_{\vec{y}_\ast},
\ee
 has not more than one eigenvalue larger than one. By using Eqs. 
% (\ref{eq2.1}), (\ref{eq3.1})-(\ref{eq3.8})
(S2), (S5)-(S12), (\ref{eq34}), (\ref{eq36}), (\ref{eq35}) we obtain the eigenvalues of $\mathcal{N}$, which read $\lambda_1 = 2^{1/3}, \lambda_2 = \cdots = \lambda_9 = 1 +  3 \, \epsilon \log 2$: given that in the non-mean-field region $\epsilon > 0$, the FP (\ref{eq33}) is unstable. 

\bibliographystyle{unsrt}
\bibliography{bibliography}

\begin{thebibliography}{10}
\bibitem{note1}
Here we chose $1/2 < \sigma < 1$ because this is the interval where the
  hierarchical Edwards-Anderson model is expected to have a finite-temperature
  phase transition \cite {franz2009overlap}.
\bibitem{note2}
We have chosen to incorporate the factor $\beta ^3 (\log 2
  /(2^{1/3}-1)^3)^{1/2}$ in the definition (\ref {eq27}) because this factor
  ensures that the FP equations for $\{ \omega _p \}$, Eqs. (\ref
  {eqf3.1})-(\ref {eqf3.8}), have integer coefficients independent of $\beta $,
  thus allowing for an exact computation of their Gr\"{o}bner basis.
\end{thebibliography}

\clearpage


\section*{Supplementary material}

# Supplemental Material for "Hierarchical spin glasses in a magnetic field: A renormalization-group study"

Michele Castellana

*Joseph Henry Laboratories of Physics and Lewis-Sigler Institute for Integrative Genomics,*
*Princeton University, Princeton, New Jersey 08544, USA*

Carlo Barbieri

*Dipartimento di Fisica, Sapienza Università di Roma, Piazzale Aldo Moro 5, 00185, Rome, Italy*

## S1. RENORMALIZATION-GROUP EQUATIONS

To obtain the recursive equations relating $s_{k+1}$, $\{r^p_{k+1}\}$, $\{w^p_{k+1}\}$ to $s_k$, $\{r^p_k\}$, $\{w^p_k\}$, we proceed as follows. We consider Eq. (8), we replace the left-hand side (LHS) with the expression for $\mathcal{Z}_{k+1}$ given by Eq. (7), and we substitute Eq. (16) into the RHS: In the resulting equation, we match the coefficients of the monomials $I^1$, $\{I_p^2\}$, $\{I_p^3\}$ in the LHS and RHS, and we obtain a set of recursive equations which relate the coefficients at the $k+1$-th step to the coefficients at the $k$-th step. This calculation has been performed with a symbolic computation tool, and it is detailed in the online Mathematica [1] notebook `symbolic_computation.nb`, which is available as an ancillary file. The resulting equations depend analytically on the number of replicas $n$, thus we could easily extract their $n \to 0$ limit, which reads

$$s_{k+1} = \frac{2s_k}{C^{1/2}} + \frac{1}{3C^{3/2}}\big[c_1\big(3w_k^2 + w_k^3 + 4w_k^4 - w_k^5 - 3w_k^6 - 4w_k^7\big) - 2c_2\big(3w_k^1 - 3w_k^2 + w_k^3 + 4w_k^4 + 2w_k^5 - 4w_k^7\big) \quad \text{(S1)}$$
$$+3c_3\big(-2w_k^1 + w_k^2 - w_k^3 + w_k^5 + w_k^6\big)\big],$$

$$r^1_{k+1} = \frac{2r^1_k}{C} - \frac{\beta^2}{2} + \frac{4}{9C^3}\big\{c_1^2\big[9\big(w_k^1\big)^2 + 6w_k^1(w_k^3 + 4w_k^4 + 2w_k^5) - 9\big(w_k^2\big)^2 - 6w_k^2(2w_k^3 + 4w_k^4 + w_k^5) - \big(w_k^3\big)^2 \quad \text{(S2)}$$
$$+8w_k^3 w_k^4 + 4w_k^3 w_k^5 - 2\big(w_k^5\big)^2\big] + 2c_1\big[c_2\big(9\big(w_k^1\big)^2 + 6w_k^1(3w_k^2 + 3w_k^3 - 4(w_k^4 + w_k^5)) - 18\big(w_k^2\big)^2 - 6w_k^2(w_k^3$$
$$-4w_k^4 - 2w_k^5) + 5\big(w_k^3\big)^2 - 8w_k^3 w_k^4 - 8w_k^3 w_k^5 + 4\big(w_k^5\big)^2\big) - c_3(3w_k^1 - 3w_k^2 + w_k^3)^2\big] - 2c_2^2(6w_k^1 - 3w_k^2 + 2w_k^3$$
$$-2w_k^5)^2\big\},$$

$$r^2_{k+1} = \frac{2r^2_k}{C} - \frac{8}{9C^3}\big\{c_1^2\big[3w_k^1\big(w_k^3 + 4w_k^4 - w_k^5 - 3w_k^6 - 8w_k^7\big) + 12w_k^7\big(w_k^2 - w_k^3 + w_k^5 + w_k^6\big) + 3w_k^2 w_k^3 \quad \text{(S3)}$$
$$+6w_k^2 w_k^5 + 9w_k^2 w_k^6 + 2(w_k^3)^2 + 8w_k^3 w_k^4 + w_k^3 w_k^5 - 3w_k^3 w_k^6 - 12w_k^4 w_k^5 - 12w_k^4 w_k^6 - 5(w_k^5)^2 - 6w_k^5 w_k^6\big]$$
$$+c_1\big[c_2\big(-9(w_k^1)^2 + 6w_k^1\big(3w_k^2 - w_k^3 - 2\big(6w_k^4 + 5w_k^5 + 3w_k^6 - 8w_k^7\big)\big) - 48w_k^7\big(w_k^2 - w_k^3 + w_k^5 + w_k^6\big) + 18w_k^2 w_k^3$$
$$+24w_k^2 w_k^4 + 12w_k^2 w_k^5 + (w_k^3)^2 - 40w_k^3 w_k^4 - 40w_k^3 w_k^5 - 24w_k^3 w_k^6 + 48w_k^4 w_k^5 + 48w_k^4 w_k^6 + 38(w_k^5)^2 + 60w_k^5 w_k^6$$
$$+18(w_k^6)^2\big) - c_3\big(3w_k^1 + 2w_k^3 - 3\big(w_k^5 + w_k^6\big)\big)\big(9w_k^1 - 6w_k^2 + 4w_k^3 - 3\big(w_k^5 + w_k^6\big)\big)\big] + c_2\big[c_2\big(-4\big(9(w_k^1)^2 + 3w_k^1$$
$$\times\big(5w_k^3 - 8w_k^4 - 11w_k^5 - 9w_k^6\big) + 3w_k^2\big(-w_k^3 + 4\big(w_k^4 + w_k^5\big) + 3w_k^6\big) + 3w_k^6\big(-5w_k^3 + 4w_k^4 + 8w_k^5\big)$$
$$+\big(w_k^3 - w_k^5\big)\big(5w_k^3 - 12w_k^4 - 14w_k^5\big) + 9(w_k^6)^2\big) - 48w_k^7\big(2w_k^1 - w_k^2 + w_k^3 - w_k^5 - w_k^6\big) + 9(w_k^2)^2\big) + 18c_3\big(-2w_k^1$$
$$+w_k^2 - w_k^3 + w_k^5 + w_k^6\big)^2\big]\big\},$$

$$r^3_{k+1} = \frac{2r^3_k}{C} - \frac{2}{9C^3}\big\{c_1^2\big[6\big(w_k^1\big(3w_k^6 + 2w_k^7 - 24w_k^8\big) + 2w_k^2\big(w_k^4 + w_k^7 + 6w_k^8\big)\big) + \big(w_k^3\big)^2 + 2w_k^3\big(2w_k^4 + 4w_k^5 \quad \text{(S4)}$$
$$+3w_k^6 + 6w_k^7 - 36w_k^8\big) + 24\big(w_k^4\big)^2 + 12w_k^4\big(w_k^5 - w_k^6 - 4w_k^7\big) + 5\big(w_k^5\big)^2 + 72w_k^8\big(w_k^5 + w_k^6\big) - 6w_k^5 w_k^6 - 48w_k^5 w_k^7$$
$$+9\big(w_k^6\big)^2 - 24w_k^6 w_k^7 + 24\big(w_k^7\big)^2\big] + 8c_2\big[c_1\big(3w_k^1\big(w_k^3 + w_k^5 - 10w_k^7 + 24w_k^8\big) + 6w_k^2 w_k^4 - 36w_k^8\big(w_k^2 + w_k^5 + w_k^6\big)$$
$$+3w_k^2 w_k^5 + 6w_k^2 w_k^7 + \big(w_k^3\big)^2 + w_k^3\big(2w_k^4 + 2w_k^5 + 3w_k^6 - 18w_k^7 + 36w_k^8\big) - 12\big(w_k^4\big)^2 - 18w_k^4 w_k^5 - 6w_k^4 w_k^6$$
$$+24w_k^4 w_k^7 - 10\big(w_k^5\big)^2 - 12w_k^5 w_k^6 + 36w_k^5 w_k^7 - 9\big(w_k^6\big)^2 + 24w_k^6 w_k^7 - 12\big(w_k^7\big)^2\big) - 3c_3\big(2w_k^1 - w_k^2 + w_k^3 - w_k^5$$
$$-w_k^6\big)\big(3w_k^1 + 2w_k^3 - 4w_k^4 - 5w_k^5 - 3w_k^6 + 4w_k^7\big)\big] + 2c_1 c_3\big[9\big(w_k^1\big)^2 + 6w_k^1\big(2w_k^3 - 8w_k^4 - 7w_k^5 - 3w_k^6 + 8w_k^7\big)$$
$$+12w_k^2\big(2w_k^4 + w_k^5 - 2w_k^7\big) + 4\big(w_k^3\big)^2 - 12w_k^3\big(2w_k^4 + 2w_k^5 + w_k^6 - 2w_k^7\big) + 3\big(w_k^5 + w_k^6\big)\big(8w_k^4 + 7w_k^5 + 3w_k^6$$
$$-8w_k^7\big)\big] - 2c_2^2\big[9\big(w_k^1\big)^2 - 6w_k^1\big(3w_k^2 + w_k^3 - 2\big(6w_k^4 + 8w_k^5 + 9w_k^6 - 18w_k^7 + 24w_k^8\big)\big) - 6w_k^2\big(w_k^3 + 2w_k^5 + 6w_k^6$$
$$-12w_k^7 + 24w_k^8\big) - 7\big(w_k^3\big)^2 + 48w_k^3 w_k^4 + 144w_k^8\big(w_k^3 - w_k^5 - w_k^6\big) + 68w_k^3 w_k^5 + 72w_k^3 w_k^6$$
$$-120w_k^3 w_k^7 - 48\big(w_k^4\big)^2 - 120w_k^4 w_k^5 - 72w_k^4 w_k^6 + 96w_k^4 w_k^7 - 88\big(w_k^5\big)^2 - 144w_k^5 w_k^6 + 192w_k^5 w_k^7 - 72\big(w_k^6\big)^2$$

$$+144w_k^6 w_k^7 - 48\left(w_k^7\right)^2\right] + 9c_3^2\left(-2w_k^1 + w_k^2 - w_k^3 + w_k^5 + w_k^6\right)^2\},$$

$$w_{k+1}^1 = \frac{2w_k^1}{C^{3/2}} + \frac{16}{27C^{9/2}}\{c_1^3\left[-27\left(w_k^1\right)^3 - 27\left(w_k^1\right)^2(w_k^3 + 8w_k^4 + 3w_k^5) + 18w_k^1(3w_k^2(w_k^3 + 4w_k^4 + w_k^5) + \left(w_k^3\right)^2 \quad \text{(S5)}$$
$$-4w_k^3 w_k^4 - 3w_k^3 w_k^5 + 2\left(w_k^5\right)^2) + (w_k^3 - w_k^5)(2w_k^5(9w_k^2 + 2w_k^5) + \left(w_k^3\right)^2 - 8w_k^3 w_k^5)\right] - 3c_1^2\left[c_2(54\left(w_k^1\right)^3\right.$$
$$+9\left(w_k^1\right)^2(3w_k^2 + 4(4w_k^3 - 4w_k^4 - 5w_k^5)) - 6w_k^1(9\left(w_k^2\right)^2 + 3w_k^2(5w_k^3 - 8(w_k^4 + w_k^5)) - 9\left(w_k^3\right)^2 + 8w_k^3 w_k^4$$
$$+20w_k^3 w_k^5 - 12\left(w_k^5\right)^2) - 18\left(w_k^2\right)^2 w_k^5 + w_k^2(-9\left(w_k^3\right)^2 + 48w_k^3 w_k^5 - 36\left(w_k^5\right)^2) + 4(w_k^3 - w_k^5)(\left(w_k^3\right)^2$$
$$-4w_k^3 w_k^5 + 2\left(w_k^5\right)^2)) - 6c_3 w_k^1(3w_k^1 - 3w_k^2 + w_k^3)^2\right] + 6c_1 c_2^2(3w_k^1 + w_k^3$$
$$-2w_k^5)(6w_k^1 - 3w_k^2 + 2w_k^3 - 2w_k^5)(9w_k^1 - 6w_k^2 + 3w_k^3 - 2w_k^5) - 4c_2^3(6w_k^1 - 3w_k^2 + 2w_k^3 - 2w_k^5)^3\},$$

$$w_{k+1}^2 = \frac{2w_k^2}{C^{3/2}} + \frac{96c_1}{C^{9/2}}\{c_1^2\left[-9\left(w_k^1\right)^2(3w_k^2 + 2w_k^3 - 2w_k^5) - 6w_k^1(3w_k^2(w_k^3 + 4w_k^4 + 2w_k^5) + 2w_k^3(w_k^3 - w_k^5)) \quad \text{(S6)}$$
$$+18\left(w_k^2\right)^3 + 18\left(w_k^2\right)^2(2w_k^3 + 4w_k^4 + w_k^5) + 3w_k^2 w_k^3(3w_k^3 - 8w_k^4 - 4w_k^5) + 2\left(w_k^3\right)^2(w_k^5 - w_k^3)\right]$$
$$+2c_1(3w_k^1 - 3w_k^2 + w_k^3)\left[2c_2(9\left(w_k^1\right)^2 - 2w_k^5(3w_k^1 - 3w_k^2 + w_k^3) + 9w_k^1 w_k^3 - 9\left(w_k^2\right)^2 - 9w_k^2 w_k^3\right.$$
$$+12w_k^2 w_k^4 + 2\left(w_k^3\right)^2) + 3c_3 w_k^2(3w_k^1 - 3w_k^2 + w_k^3)\right] + 4c_2^2(3w_k^1 - 3w_k^2 + w_k^3)^2(-6w_k^1 + 3w_k^2 - 2w_k^3 + 2w_k^5)\},$$

$$w_{k+1}^3 = \frac{2w_k^3}{C^{3/2}} + \frac{16}{9C^{9/2}}\{c_1^3\left[9\left(w_k^1\right)^2(3w_k^2 - w_k^3 - 8w_k^4 - 8w_k^5 - 3w_k^6 + 8w_k^7) + 3w_k^1(6w_k^2(w_k^3 + 4w_k^4 - 3w_k^6 \quad \text{(S7)}$$
$$-8w_k^7) - 3\left(w_k^3\right)^2 - 2w_k^3(24w_k^4 + 18w_k^5 + 9w_k^6 - 8w_k^7) + 24w_k^6(w_k^4 + w_k^5) + 4w_k^5(6w_k^4 + 5w_k^5))$$
$$+18\left(w_k^2\right)^2(w_k^3 + 2w_k^5 + 3w_k^6 + 4w_k^7) + 3w_k^2(11\left(w_k^3\right)^2 + 2w_k^3(20w_k^4 + 8w_k^5 + 3w_k^6 - 8w_k^7) - 2(12w_k^4(w_k^5$$
$$+w_k^6) + w_k^5(5w_k^5 + 6w_k^6))) + 5\left(w_k^3\right)^3 - 40\left(w_k^3\right)^2 w_k^4 - 28\left(w_k^3\right)^2 w_k^5 - 15\left(w_k^3\right)^2 w_k^6 + 8\left(w_k^3\right)^2 w_k^7$$
$$+24w_k^3 w_k^4 w_k^5 + 24w_k^3 w_k^4 w_k^6 + 24w_k^3\left(w_k^5\right)^2 + 24w_k^3 w_k^5 w_k^6 - 8\left(w_k^5\right)^3 - 12\left(w_k^5\right)^2 w_k^6\right] + c_1^2\left[c_2\left(-216\left(w_k^1\right)^3\right.\right.$$
$$+9\left(w_k^1\right)^2(3w_k^2 - 42w_k^3 + 64w_k^4 + 88w_k^5 + 60w_k^6 - 32w_k^7) + 6w_k^1(18\left(w_k^2\right)^2 - 3w_k^2(w_k^3 + 8(6w_k^4 + 5w_k^5 + 3w_k^6$$
$$-4w_k^7)) - 2(25\left(w_k^3\right)^2 - 2w_k^3(26w_k^4 + 34w_k^5 + 21w_k^6 - 8w_k^7) + 6(4w_k^4(w_k^5 + w_k^6) + w_k^5(5w_k^5 + 6w_k^6))))$$
$$+18\left(w_k^2\right)^2(10w_k^3 + 16w_k^4 + 8w_k^5 + 3w_k^6 - 16w_k^7) + 3w_k^2(37\left(w_k^3\right)^2 - 8w_k^3(22w_k^4 + 19w_k^5 + 12w_k^6 - 8w_k^7)$$
$$+12(8w_k^4(w_k^5 + w_k^6) + w_k^5(5w_k^5 + 6w_k^6))) - 2(33\left(w_k^3\right)^3 - 2\left(w_k^3\right)^2(36w_k^4 + 46w_k^5 + 27w_k^6 - 8w_k^7)$$
$$+24w_k^3(2w_k^4 + 3w_k^5)(w_k^5 + w_k^6) - 12\left(w_k^5\right)^2(2w_k^5 + 3w_k^6))) + 2c_3(3w_k^1 - 3w_k^2 + w_k^3)^2(6w_k^1 + 7w_k^3 - 6(w_k^5$$
$$+w_k^6))\right] + 2c_1 c_2\left[c_2(486\left(w_k^1\right)^3 - 9\left(w_k^1\right)^2(57w_k^2 - 96w_k^3 + 48w_k^4 + 104w_k^5 + 78w_k^6 - 16w_k^7) + 6w_k^1(9\left(w_k^2\right)^2\right.$$
$$+3w_k^2(-45w_k^3 + 40w_k^4 + 56w_k^5 + 42w_k^6 - 16w_k^7) + 73\left(w_k^3\right)^2 - 2w_k^3(28w_k^4 + 70w_k^5 + 45w_k^6 - 8w_k^7) + 12(2w_k^4$$
$$\times(w_k^5 + w_k^6) + w_k^5(5w_k^5 + 6w_k^6))) + 54\left(w_k^2\right)^3 + 18\left(w_k^2\right)^2(9w_k^3 - 4(4w_k^4 + 4w_k^5 + 3w_k^6 - 2w_k^7)) - 3w_k^2$$
$$\times(79\left(w_k^3\right)^2 - 4w_k^3(24w_k^4 + 40w_k^5 + 27w_k^6 - 8w_k^7) + 12(4w_k^4(w_k^5 + w_k^6) + w_k^5(5w_k^5 + 6w_k^6))) + 2(34\left(w_k^3\right)^3$$
$$-\left(w_k^3\right)^2(32w_k^4 + 88w_k^5 + 51w_k^6 - 8w_k^7) + 24w_k^3(w_k^4 + 3w_k^5)(w_k^5 + w_k^6) - 12\left(w_k^5\right)^2(2w_k^5 + 3w_k^6))) - 12c_3$$
$$\times(3w_k^1 - 3w_k^2 + w_k^3)^2(2w_k^1 - w_k^2 + w_k^3 - w_k^5 - w_k^6)\right] + 4c_2^3(6w_k^1 - 3w_k^2 + 2w_k^3 - 2w_k^5)^2(-6w_k^1 + 3w_k^2$$
$$-4w_k^3 + 4w_k^5 + 6w_k^6)\},$$

$$w_{k+1}^4 = \frac{2w_k^4}{C^{3/2}} + \frac{24}{27C^{9/2}}\{\left[\left(w_k^3\right)^3 + 2(10w_k^4 + 3(w_k^5 + w_k^6 - 3w_k^7 + 4w_k^8))\left(w_k^3\right)^2 + 2\left(-36\left(w_k^4\right)^2 - 4(7w_k^5\right.\right. \quad \text{(S8)}$$
$$+3w_k^6 - 4w_k^7)w_k^4 + w_k^5(8w_k^7 - 7w_k^5) + 6w_k^2(4w_k^4 + w_k^5 + 3w_k^7 - 12w_k^8))w_k^3 + 9\left(w_k^1\right)^2(w_k^3 - 4w_k^4 + 4w_k^5$$
$$-2w_k^7 + 24w_k^8) + 4(9(w_k^4 + w_k^7 + 6w_k^8)\left(w_k^2\right)^2 + 3(14\left(w_k^4\right)^2 + 6w_k^5 w_k^4 + \left(w_k^5\right)^2 - 2(4w_k^4 + w_k^5)w_k^7)w_k^2$$
$$+2(3(w_k^5 + w_k^6)\left(w_k^4\right)^2 + \left(w_k^5\right)^2 w_k^4 + \left(w_k^5\right)^2(w_k^5 - w_k^7))) + 3w_k^1(\left(w_k^3\right)^2 + 4(w_k^5 - 5w_k^7 + 12w_k^8)w_k^3$$
$$-2(32\left(w_k^4\right)^2 + 16w_k^5 w_k^4 + 7\left(w_k^5\right)^2 - 8(2w_k^4 + w_k^5)w_k^7 + 6w_k^2(w_k^7 + 12w_k^8)))\right]c_1^3 + \left[c_2(81\left(w_k^1\right)^3 + 9(5w_k^3\right.$$
$$-4(7w_k^4 + 8w_k^5 - 14w_k^7 + 24w_k^8))\left(w_k^1\right)^2 + 3(9\left(w_k^3\right)^2 - 8(23w_k^4 + 12w_k^5 + 3w_k^6 - 22w_k^7 + 24w_k^8)w_k^3$$
$$+4(56\left(w_k^4\right)^2 + 52w_k^5 w_k^4 + 12w_k^6 w_k^4 + 23\left(w_k^5\right)^2 - 8(4w_k^4 + 3w_k^5)w_k^7) + 12w_k^2(w_k^3 - 2(w_k^4 + 8w_k^7$$
$$-24w_k^8)))w_k^1 + 15\left(w_k^3\right)^3 - 48\left(w_k^5\right)^3 + 256w_k^3\left(w_k^4\right)^2 + 92w_k^3\left(w_k^5\right)^2 - 48w_k^4\left(w_k^5\right)^2 - 188\left(w_k^3\right)^2 w_k^4$$
$$-88\left(w_k^3\right)^2 w_k^5 - 96\left(w_k^4\right)^2 w_k^5 + 304w_k^3 w_k^4 w_k^5 - 48\left(w_k^3\right)^2 w_k^6 - 96\left(w_k^4\right)^2 w_k^6 + 144w_k^3 w_k^4 w_k^6 + 120\left(w_k^3\right)^2 w_k^7$$
$$+48\left(w_k^5\right)^2 w_k^7 - 128w_k^3 w_k^4 w_k^7 - 96w_k^3 w_k^5 w_k^7 + 36\left(w_k^2\right)^2(6w_k^4 + 2w_k^5 + 5w_k^7 - 24w_k^8) - 96\left(w_k^3\right)^2 w_k^8$$
$$+12w_k^2(\left(w_k^3\right)^2 + (22w_k^4 + 8w_k^5 + 6w_k^6 - 32w_k^7 + 48w_k^8)w_k^3 - 4(12\left(w_k^4\right)^2 + (10w_k^5 + 3w_k^6 - 8w_k^7)w_k^4$$

$$+w_k^5(2w_k^5 - 3w_k^7)))) - 4c_3(3w_k^1 - 3w_k^2 + w_k^3)(2(w_k^3)^2 - (11w_k^4 + 5w_k^5 + 3w_k^6 - 4w_k^7)w_k^3 + 6w_k^4(w_k^5 + w_k^6)$$
$$+3w_k^2(7w_k^4 + 2w_k^5 - 4w_k^7) + 3w_k^1(w_k^3 - 9w_k^4 - 2w_k^5 + 4w_k^7))]c_1^2 - 2\big[(135(w_k^1)^3 + 9(23w_k^3 - 4(15w_k^4$$
$$+15w_k^5 + 3w_k^6 - 17w_k^7 + 12w_k^8))(w_k^1)^2 - 3(18(w_k^2)^2 + 12(2w_k^3 - 14w_k^4 - 11w_k^5 - 6w_k^6 + 22w_k^7 - 24w_k^8)w_k^2$$
$$-47(w_k^3)^2 - 4(24(w_k^4)^2 + 44w_k^5 w_k^4 + 12w_k^6 w_k^4 + 25(w_k^5)^2 - 8(2w_k^4 + 3w_k^5)w_k^7) + 8w_k^3(25w_k^4 + 20w_k^5$$
$$+6w_k^6 - 21w_k^7 + 12w_k^8))w_k^1 + 33(w_k^3)^3 - 48(w_k^5)^3 + 112w_k^3(w_k^4)^2 + 100w_k^3(w_k^5)^2 - 48w_k^4(w_k^5)^2$$
$$-156(w_k^3)^2 w_k^4 - 112(w_k^3)^2 w_k^5 - 48(w_k^4)^2 w_k^5 + 224w_k^3 w_k^4 w_k^5 - 48(w_k^3)^2 w_k^6 - 48(w_k^4)^2 w_k^6 + 96w_k^3 w_k^4 w_k^6$$
$$+100(w_k^3)^2 w_k^7 + 48(w_k^5)^2 w_k^7 - 64w_k^3 w_k^4 w_k^7 - 96w_k^3 w_k^5 w_k^7 - 48(w_k^3)^2 w_k^8 - 18(w_k^2)^2(w_k^3 + 4w_k^4 + 4w_k^5$$
$$+6w_k^6 - 16w_k^7 + 24w_k^8) - 12w_k^2(5(w_k^3)^2 - 3(10w_k^4 + 7w_k^5 + 4w_k^6 - 10w_k^7 + 8w_k^8)w_k^3 + 2(10(w_k^4)^2$$
$$+2(8w_k^5 + 3w_k^6 - 4w_k^7)w_k^4 + w_k^5(5w_k^5 - 6w_k^7))))c_2^2 - 4c_3(3w_k^1 - 3w_k^2 + w_k^3)(9(w_k^1)^2 - 3(3w_k^2 - 5w_k^3$$
$$+8w_k^4 + 5w_k^5 + 3w_k^6 - 4w_k^7)w_k^1 + 5(w_k^3)^2 - 10w_k^3 w_k^4 - 8w_k^3 w_k^5 + 6w_k^4 w_k^5 - 6w_k^3 w_k^6 + 6w_k^4 w_k^6 + 3w_k^2(-3w_k^3$$
$$+6w_k^4 + 5w_k^5 + 3w_k^6 - 4w_k^7) + 4w_k^3 w_k^7)c_2 + 3c_3^2(3w_k^1 - 3w_k^2 + w_k^3)^2(2w_k^1 - w_k^2 + w_k^3 - w_k^5 - w_k^6)\big]c_1$$
$$+4c_2^3(6w_k^1 - 3w_k^2 + 2w_k^3 - 2w_k^5)^2[3w_k^1 + w_k^3 - 4(w_k^4 + w_k^5 - w_k^7)]\Big\},$$

$$w_{k+1}^5 = \frac{2w_k^5}{C^{3/2}} + \frac{1}{9C^{9/2}}\Big\{16\big[3(w_k^3)^3 + 16w_k^4(w_k^3)^2 + 14w_k^5(w_k^3)^2 + 9w_k^6(w_k^3)^2 - 46(w_k^5)^2 w_k^3 \qquad (S9)$$
$$-18(w_k^6)^2 w_k^3 - 80w_k^4 w_k^5 w_k^3 - 24w_k^4 w_k^6 w_k^3 - 48w_k^5 w_k^6 w_k^3 + 22(w_k^5)^3 + 48w_k^4(w_k^5)^2 + 18w_k^5(w_k^6)^2$$
$$+36(w_k^5)^2 w_k^6 + 48w_k^4 w_k^5 w_k^6 + 9(w_k^1)^2(w_k^3 + 8w_k^4 - 3w_k^6 - 16w_k^7) - 8(3(w_k^3)^2 - (7w_k^5 + 3w_k^6)w_k^3$$
$$+3w_k^5(w_k^5 + w_k^6))w_k^7 + 6w_k^1(2(w_k^3)^2 + (12w_k^4 + 4w_k^5 + 3w_k^6 - 20w_k^7)w_k^3 - 13(w_k^5)^2 - 9(w_k^6)^2 - 12w_k^5 w_k^6$$
$$-4w_k^4(8w_k^5 + 3w_k^6) + 12(2w_k^5 + w_k^6)w_k^7 + 3w_k^2(w_k^5 + 3w_k^6 + 6w_k^7)) + 3w_k^2((w_k^3)^2 + 2(4w_k^5 + 3w_k^6 + 8w_k^7)w_k^3$$
$$+24w_k^4 w_k^5 + 9((w_k^5)^2 + (w_k^6)^2) - 4(7w_k^5 + 3w_k^6)w_k^7)\big]c_1^3 - 16\big[c_3(162(w_k^1)^3 - 9(15w_k^2 - 26w_k^3 + 40w_k^5$$
$$+24w_k^6)(w_k^1)^2 + 6(18(w_k^3)^2 - 2(26w_k^5 + 15w_k^6)w_k^3 + 9(w_k^5 + w_k^6)(3w_k^5 + w_k^6) + 3w_k^2(-8w_k^3 + 19w_k^5$$
$$+9w_k^6))w_k^1 - 54(w_k^2)^2 w_k^5 + 2(8(w_k^3)^3 - 3(11w_k^5 + 6w_k^6)(w_k^3)^2 + 3(w_k^5 + w_k^6)(11w_k^5 + 3w_k^6)w_k^3$$
$$-9w_k^5(w_k^5 + w_k^6)^2) - 3w_k^2(12(w_k^3)^2 - 4(13w_k^5 + 6w_k^6)w_k^3 + 3(w_k^5 + w_k^6)(11w_k^5 + 3w_k^6))) - 2c_2(11(w_k^3)^3$$
$$-64w_k^4(w_k^3)^2 - 114w_k^5(w_k^3)^2 - 72w_k^6(w_k^3)^2 + 194(w_k^5)^2 w_k^3 + 72(w_k^6)^2 w_k^3 + 208w_k^4 w_k^5 w_k^3 + 72w_k^4 w_k^6 w_k^3$$
$$+234w_k^5 w_k^6 w_k^3 - 80(w_k^5)^3 - 120w_k^4(w_k^5)^2 - 72w_k^5(w_k^6)^2 - 144(w_k^5)^2 w_k^6 - 120w_k^4 w_k^5 w_k^6 + 8(9(w_k^3)^2$$
$$-(20w_k^5 + 9w_k^6)w_k^3 + 9w_k^5(w_k^5 + w_k^6))w_k^7 + 18(w_k^2)^2(w_k^5 + 2w_k^7) + 9(w_k^1)^2(3w_k^2 + 3w_k^3 - 40w_k^4 - 36w_k^5$$
$$-24w_k^6 + 48w_k^7) + 3w_k^1(3w_k^2(7w_k^3 + 20w_k^4 + 15w_k^5 + 9w_k^6 - 40w_k^7) + 2(7(w_k^3)^2 - (52w_k^4 + 67w_k^5 + 48w_k^6$$
$$-60w_k^7)w_k^3 + 63(w_k^5)^2 + 36(w_k^6)^2 + 84w_k^4 w_k^5 + 36w_k^4 w_k^6 + 81w_k^5 w_k^6 - 68w_k^5 w_k^7 - 36w_k^6 w_k^7))$$
$$+3w_k^2(6(w_k^3)^2 + (24w_k^4 + 27w_k^5 + 21w_k^6 - 52w_k^7)w_k^3 - 4(13(w_k^5)^2 + 17w_k^4 w_k^5 + 15w_k^6 w_k^5 - 19w_k^7 w_k^5$$
$$+9(w_k^6)^2 + 3w_k^4 w_k^6 - 9w_k^6 w_k^7)))\big]c_1^2 - 32c_2\big[c_2(270(w_k^1)^3 - 9(9w_k^2 - 68w_k^3 + 96w_k^4 + 140w_k^5 + 114w_k^6$$
$$-96w_k^7)(w_k^1)^2 - 6(9(w_k^2)^2 + 3(13w_k^3 - 36w_k^4 - 46w_k^5 - 42w_k^6 + 44w_k^7)w_k^2 - 63(w_k^3)^2 - 2(83(w_k^5)^2$$
$$+8(15w_k^6 - 8w_k^7)w_k^5 + 36w_k^4(2w_k^5 + w_k^6) + 9w_k^6(5w_k^6 - 4w_k^7)) + 6w_k^3(20w_k^4 + 38w_k^5 + 29w_k^6 - 20w_k^7))w_k^1$$
$$-18(w_k^2)^2(w_k^3 + 4w_k^4 + 4w_k^5 + 6w_k^6 - 8w_k^7) + 2(34(w_k^3)^3 - (72w_k^4 + 172w_k^5 + 117w_k^6 - 72w_k^7)$$
$$\times(w_k^3)^2 + 2(119(w_k^5)^2 + 88w_k^4 w_k^5 + 156w_k^6 w_k^5 - 76w_k^7 w_k^5 + 45(w_k^6)^2 + 36w_k^4 w_k^6 - 36w_k^6 w_k^7)w_k^3$$
$$-2w_k^5(47(w_k^5)^2 + 90w_k^6 w_k^5 + 45(w_k^6)^2 + 48w_k^4(w_k^5 + w_k^6)) + 72w_k^5(w_k^5 + w_k^6)w_k^7) - 3w_k^2(29(w_k^3)^2$$
$$-4(22w_k^4 + 40w_k^5 + 33w_k^6 - 28w_k^7)w_k^3 + 2(73(w_k^5)^2 + 108w_k^6 w_k^5 - 68w_k^7 w_k^5 + 45(w_k^6)^2 + 8w_k^4(8w_k^5$$
$$+3w_k^6) - 36w_k^6 w_k^7))) - 6c_3(2w_k^1 - w_k^2 + w_k^3 - w_k^5 - w_k^6)(36(w_k^1)^2 - 3(9w_k^2 - 10w_k^3 + 12w_k^5 + 6w_k^6)w_k^1$$
$$+6(w_k^3)^2 + 6(w_k^5)^2 - 14w_k^3 w_k^5 - 6w_k^3 w_k^6 + 6w_k^5 w_k^6 + 3w_k^2(-4w_k^3 + 7w_k^5 + 3w_k^6))\big]c_1 + 3\big[64c_2^2(6w_k^1$$
$$-3w_k^2 + 2w_k^3 - 2w_k^5)(2w_k^1 - w_k^2 + w_k^3 - w_k^5 - w_k^6)(3c_3(-2w_k^1 + w_k^2 - w_k^3 + w_k^5 + w_k^6) + 2c_2(3w_k^1 + 3w_k^3$$
$$-4w_k^4 - 6(w_k^5 + w_k^6) + 4w_k^7))\big]\Big\},$$

$$w_k^6 = \frac{2w_k^6}{C^{3/2}} - \frac{16}{27C^{9/2}}\Big\{\big[216w_k^7(w_k^1)^2 - 9\big((w_k^3)^2 + 4(2w_k^4 - 3w_k^6 - 10w_k^7)w_k^3 + 6w_k^2(w_k^5 + 2w_k^7) - 4(2w_k^4 \qquad (S10)$$
$$\times(w_k^5 + 6w_k^6) + 3((w_k^5)^2 + 3w_k^6 w_k^5 - 4w_k^7 w_k^5 - 6w_k^6 w_k^7))\big)w_k^1 + 2\big(-32(w_k^5)^3 + 9(5w_k^3 - 4w_k^4)(w_k^5)^2$$

$$
\begin{aligned}
&-6w_k^3(w_k^3 - 10w_k^4)w_k^5 - 27\left(w_k^6\right)^3 + 27(w_k^3 - 4(w_k^4 + w_k^5))\left(w_k^6\right)^2 - 4\left(w_k^3\right)^2(w_k^3 + 6w_k^4) + 9(w_k^3 - w_k^5) \\
&\times (w_k^3 + 16w_k^4 + 13w_k^5)w_k^6 + 12(5\left(w_k^3\right)^2 - (11w_k^5 + 15w_k^6)w_k^3 + 3(w_k^5 + w_k^6)(2w_k^5 + 3w_k^6))w_k^7) - 9w_k^2 \\
&\times \left(\left(w_k^3\right)^2 + 4(w_k^5 + 3w_k^6 + 4w_k^7)w_k^3 + \left(w_k^5\right)^2 + 24w_k^4 w_k^6 + 4w_k^5(3w_k^6 - 5w_k^7) - 9w_k^6(w_k^6 + 4w_k^7)\right)\big]c_1^3 \\
&-3\big[3c_3(18\left(w_k^6\right)^3 + 3(-30w_k^1 + 15w_k^2 + 16(w_k^5 - w_k^3))\left(w_k^6\right)^2 + 2(54\left(w_k^1\right)^2 + (-45w_k^2 + 66w_k^3 - 72w_k^5)w_k^1 \\
&+ 9\left(w_k^2\right)^2 + (19w_k^3 - 21w_k^5)(w_k^3 - w_k^5) + w_k^2(33w_k^5 - 30w_k^3))w_k^6 - (3w_k^1 + 2w_k^3 - 3w_k^5)(6\left(w_k^1\right)^2 \\
&+ (-3w_k^2 + 12w_k^3 - 14w_k^5)w_k^1 + 4(w_k^3 - w_k^5)^2 + w_k^2(7w_k^5 - 6w_k^3))) + c_2(9(3w_k^2 - 4(4w_k^4 + 8w_k^5 + 9w_k^6 \\
&- 16w_k^7))\left(w_k^1\right)^2 + 6(3w_k^2(2w_k^3 + 4w_k^4 + 5w_k^5 + 3w_k^6 - 24w_k^7) - 2(\left(w_k^3\right)^2 + (32w_k^4 + 39w_k^5 + 36w_k^6 - 68w_k^7) \\
&\times w_k^3 - 3(16\left(w_k^5\right)^2 + 12w_k^4 w_k^5 + 39w_k^6 w_k^5 + 15\left(w_k^6\right)^2 + 28w_k^4 w_k^6) + 4(19w_k^5 + 27w_k^6)w_k^7))w_k^1 + 3w_k^2(15 \\
&\times \left(w_k^3\right)^2 + 2(24w_k^4 + 17w_k^5 + 3w_k^6 - 60w_k^7)w_k^3 - 4(16\left(w_k^5\right)^2 + (42w_k^6 - 34w_k^7)w_k^5 + 14w_k^4(w_k^5 + 3w_k^6) \\
&+ 9w_k^6(w_k^6 - 6w_k^7))) + 18\left(w_k^2\right)^2(3w_k^6 + 4w_k^7) + 4(\left(w_k^3\right)^3 - (40w_k^4 + 47w_k^5 + 45w_k^6 - 64w_k^7)\left(w_k^3\right)^2 \\
&+ (98\left(w_k^5\right)^2 + 88w_k^4 w_k^5 + 225w_k^6 w_k^5 + 99\left(w_k^6\right)^2 + 156w_k^4 w_k^6 - 4(34w_k^5 + 45w_k^6)w_k^7)w_k^3 - 2(26\left(w_k^5\right)^3 \\
&+ 90w_k^6\left(w_k^5\right)^2 + 90\left(w_k^6\right)^2 w_k^5 + 27\left(w_k^6\right)^3 + 6w_k^4(w_k^5 + w_k^6)(4w_k^5 + 9w_k^6)) + 36(w_k^5 + w_k^6)(2w_k^5 + 3w_k^6)w_k^7))\big] \\
&\times c_1^2 + 12c_2\big[c_2(54\left(w_k^1\right)^3 - 9(3w_k^2 - 14w_k^3 + 24w_k^4 + 46w_k^5 + 48w_k^6 - 40w_k^7)\left(w_k^1\right)^2 + 6(19\left(w_k^3\right)^2 - 2(28w_k^4 \\
&+ 49w_k^5 + 51w_k^6 - 38w_k^7)w_k^3 + w_k^2(-3w_k^3 + 30w_k^4 + 45w_k^5 + 36w_k^6 - 54w_k^7) + 2(41\left(w_k^5\right)^2 + 30w_k^4 w_k^5 \\
&+ 93w_k^6 w_k^5 - 40w_k^7 w_k^5 + 45\left(w_k^6\right)^2 + 48w_k^4 w_k^6 - 54w_k^6 w_k^7))w_k^1 - 18\left(w_k^2\right)^2(w_k^3 + 2(w_k^4 + w_k^5 - 2w_k^7)) - 3w_k^2 \\
&\times (6\left(w_k^3\right)^2 - 4(13w_k^4 + 17w_k^5 + 15w_k^6 - 18w_k^7)w_k^3 + 65\left(w_k^5\right)^2 + 63\left(w_k^6\right)^2 + 56w_k^4 w_k^5 + 96w_k^4 w_k^6 + 144w_k^5 w_k^6 \\
&- 76w_k^5 w_k^7 - 108w_k^6 w_k^7) + 2(16\left(w_k^3\right)^3 - (56w_k^4 + 101w_k^5 + 105w_k^6 - 68w_k^7)\left(w_k^3\right)^2 + (157\left(w_k^5\right)^2 \\
&+ 348w_k^6 w_k^5 + 171\left(w_k^6\right)^2 + 4w_k^4(29w_k^5 + 42w_k^6) - 20(7w_k^5 + 9w_k^6)w_k^7)w_k^3 - 3(w_k^5 + w_k^6)(4w_k^4(5w_k^5 \\
&+ 9w_k^6) + 3(8\left(w_k^5\right)^2 + 19w_k^6 w_k^5 - 8w_k^7 w_k^5 + 9\left(w_k^6\right)^2 - 12w_k^6 w_k^7)))) - 3c_3(2w_k^1 - w_k^2 + w_k^3 - w_k^5 - w_k^6) \\
&\times (18\left(w_k^1\right)^2 - 3(3w_k^2 + 2(-5w_k^3 + 6w_k^5 + 9w_k^6))w_k^1 + 3w_k^2(-4w_k^3 + 5w_k^5 + 9w_k^6) + 2(5\left(w_k^3\right)^2 \\
&- (11w_k^5 + 15w_k^6)w_k^3 + 3(w_k^5 + w_k^6)(2w_k^5 + 3w_k^6)))\big]c_1 - 4c_2^2(6w_k^1 - 3w_k^2 + 4w_k^3 - 4w_k^5 - 6w_k^6)\big[c_2 \\
&\times (72\left(w_k^1\right)^2 - 6(3w_k^2 - 19w_k^3 + 24w_k^4 + 37w_k^5 + 33w_k^6 - 24w_k^7)w_k^1 - 9\left(w_k^2\right)^2 + 38\left(w_k^3\right)^2 + 92\left(w_k^5\right)^2 \\
&+ 72\left(w_k^6\right)^2 - 30w_k^2 w_k^3 + 72w_k^2 w_k^4 - 72w_k^3 w_k^4 + 84w_k^2 w_k^5 - 130w_k^3 w_k^5 + 72w_k^4 w_k^5 + 72w_k^2 w_k^6 - 114w_k^3 w_k^6 \\
&+ 72w_k^4 w_k^6 + 168w_k^5 w_k^6 - 72(w_k^2 - w_k^3 + w_k^5 + w_k^6)w_k^7) - 27c_3(-2w_k^1 + w_k^2 - w_k^3 + w_k^5 + w_k^6)^2\big]\Big\},
\end{aligned}
$$

$$
w_{k+1}^7 = \frac{2w_k^7}{C^{3/2}} + \frac{8}{9C^{9/2}}\Big\{\big[\left(w_k^3\right)^3 + (16w_k^4 + 18w_k^5 + 15w_k^6 + 58w_k^7 - 192w_k^8)\left(w_k^3\right)^2 + (96\left(w_k^4\right)^2 + 104w_k^5 w_k^4 \tag{S11}
$$

$$
\begin{aligned}
&+ 24w_k^6 w_k^4 - 304w_k^7 w_k^4 + 45\left(w_k^5\right)^2 + 45\left(w_k^6\right)^2 + 168\left(w_k^7\right)^2 + 48w_k^5 w_k^6 - 372w_k^5 w_k^7 - 252w_k^6 w_k^7 \\
&+ 6w_k^1(4w_k^4 + 5w_k^5 + 9w_k^6 + 24w_k^7 - 120w_k^8) + 432(w_k^5 + w_k^6)w_k^8 + 12w_k^2(2w_k^4 + 3w_k^7 + 24w_k^8))w_k^3 \\
&+ 9\left(w_k^1\right)^2(3w_k^6 + 10w_k^7 - 72w_k^8) + 3w_k^1(48\left(w_k^4\right)^2 + 8(5w_k^5 - 24w_k^7)w_k^4 + 19\left(w_k^5\right)^2 + 9\left(w_k^6\right)^2 + 112 \\
&\times \left(w_k^7\right)^2 + 12w_k^5 w_k^6 + 12w_k^2 w_k^7 - 204w_k^5 w_k^7 - 132w_k^6 w_k^7 + 144(w_k^2 + 2(w_k^5 + w_k^6))w_k^8) + 2\big(-36\left(w_k^5\right)^3 \\
&- 78w_k^6\left(w_k^5\right)^2 + 158w_k^7\left(w_k^5\right)^2 - 63\left(w_k^6\right)^2 w_k^5 - 84\left(w_k^7\right)^2 w_k^5 + 252w_k^6 w_k^7 w_k^5 - 27\left(w_k^6\right)^3 - 84w_k^6\left(w_k^7\right)^2 \\
&- 72\left(w_k^4\right)^2(w_k^5 + w_k^6) + 90\left(w_k^6\right)^2 w_k^7 + 4w_k^4\big(-23\left(w_k^5\right)^2 - 33w_k^6 w_k^5 - 9\left(w_k^6\right)^2 + 42(w_k^5 + w_k^6)w_k^7\big) \\
&- 108(w_k^5 + w_k^6)^2 w_k^8 + 3w_k^2(3\left(w_k^5\right)^2 + 6w_k^6 w_k^5 + 28w_k^7 w_k^5 - 28\left(w_k^7\right)^2 + 12w_k^6 w_k^7 + 4w_k^4(2w_k^5 + 3w_k^6 \\
&+ 10w_k^7) - 72(w_k^5 + w_k^6)w_k^8))\big]c_1^3 + \big[c_3((3w_k^1 + 2w_k^3 - 3(w_k^5 + w_k^6))(27\left(w_k^1\right)^2 + 3(13w_k^3 - 76w_k^4 - 65w_k^5 \\
&- 33w_k^6)w_k^1 + 14\left(w_k^3\right)^2 - w_k^3(104w_k^4 + 101w_k^5 + 57w_k^6) + 12((w_k^5 + w_k^6)(7w_k^4 + 6w_k^5 + 3w_k^6) + w_k^2(12w_k^4 \\
&+ 8w_k^5 + 3w_k^6))) + 4(216\left(w_k^1\right)^2 - 12(15w_k^2 - 19w_k^3 + 21(w_k^5 + w_k^6))w_k^1 + 27\left(w_k^2\right)^2 + 59\left(w_k^3\right)^2 \\
&+ 63\left(w_k^5\right)^2 + 63\left(w_k^6\right)^2 - 102w_k^2 w_k^3 + 126w_k^2 w_k^5 - 126w_k^3 w_k^5 + 126(w_k^2 - w_k^3 + w_k^5)w_k^6)w_k^7) + c_2(19 \\
&\times \left(w_k^3\right)^3 + (188w_k^4 + 193w_k^5 + 177w_k^6 - 832w_k^7 + 1200w_k^8)\left(w_k^3\right)^2 + 4\big(-188\left(w_k^4\right)^2 + (-426w_k^5 - 282w_k^6 \\
&+ 472w_k^7)w_k^4 - 261\left(w_k^5\right)^2 - 189\left(w_k^6\right)^2 - 252\left(w_k^7\right)^2 - 396w_k^5 w_k^6 + 814w_k^5 w_k^7 + 630w_k^6 w_k^7 + 3w_k^2(18w_k^4 \\
&+ 10w_k^5 + 3w_k^6 + 46w_k^7 - 168w_k^8) - 648(w_k^5 + w_k^6)w_k^8\big)w_k^3 + 9\left(w_k^1\right)^2(7w_k^3 + 28w_k^4 + 37w_k^5 + 21w_k^6 \\
&- 232w_k^7 + 480w_k^8) + 6w_k^1(13\left(w_k^3\right)^2 + (72w_k^4 + 81w_k^5 + 57w_k^6 - 444w_k^7 + 768w_k^8)w_k^3 - 2(112\left(w_k^4\right)^2
\end{aligned}
$$

$$
\begin{aligned}
&+2(115w_k^5+75w_k^6-152w_k^7)w_k^4+133\left(w_k^5\right)^2+2w_k^5(96w_k^6-239w_k^7)+3(27w_k^6-14w_k^7)(w_k^6-4w_k^7))\\
&+6w_k^2(6w_k^4+4w_k^5+3w_k^6+22w_k^7-96w_k^8)-864(w_k^5+w_k^6)w_k^8)+4\left(209\left(w_k^5\right)^3+495w_k^6\left(w_k^5\right)^2\right.\\
&-600w_k^7\left(w_k^5\right)^2+405\left(w_k^6\right)^2w_k^5+252\left(w_k^7\right)^2w_k^5-1008w_k^6w_k^7w_k^5+135\left(w_k^6\right)^3+252w_k^6\left(w_k^7\right)^2+228\left(w_k^4\right)^2\\
&\times(w_k^5+w_k^6)-396\left(w_k^6\right)^2w_k^7+6w_k^4(69\left(w_k^5\right)^2+110w_k^6w_k^5+39\left(w_k^6\right)^2-84(w_k^5+w_k^6)w_k^7)+324(w_k^5+w_k^6)^2\\
&\times w_k^8+9\left(w_k^2\right)^2\left(w_k^7+12w_k^8\right)+3w_k^2(36\left(w_k^4\right)^2+4(13w_k^5+6w_k^6-34w_k^7)w_k^4+23\left(w_k^5\right)^2+18\left(w_k^6\right)^2+84\\
&\times\left(w_k^7\right)^2+30w_k^5w_k^6-172w_k^5w_k^7-120w_k^6w_k^7+216(w_k^5+w_k^6)w_k^8)))\bigg]c_1^2+2\bigg[\left(81\left(w_k^1\right)^3+9(9w_k^2\right.\\
&+25w_k^3-2(86w_k^4+107w_k^5+90w_k^6-218w_k^7+264w_k^8))\left(w_k^1\right)^2+3\left(71\left(w_k^3\right)^2-8(87w_k^4+110w_k^5+93w_k^6\right.\\
&-194w_k^7+204w_k^8)w_k^3+2\left(288\left(w_k^4\right)^2+8(103w_k^5+78w_k^6-80w_k^7)w_k^4+597\left(w_k^5\right)^2+405\left(w_k^6\right)^2+336\left(w_k^7\right)^2\right.\\
&+960w_k^5w_k^6-1300w_k^5w_k^7-1068w_k^6w_k^7+864(w_k^5+w_k^6)w_k^8)+6w_k^2(7w_k^3+6(4w_k^4+5w_k^5+5w_k^6-22w_k^7\\
&+40w_k^8)))w_k^1+65\left(w_k^3\right)^3-1348\left(w_k^5\right)^3-864\left(w_k^6\right)^3+896w_k^3\left(w_k^4\right)^2+2178w_k^3\left(w_k^5\right)^2-2208w_k^4\left(w_k^5\right)^2\\
&+1566w_k^3\left(w_k^6\right)^2-1440w_k^4\left(w_k^6\right)^2-2808w_k^5\left(w_k^6\right)^2+1008w_k^3\left(w_k^7\right)^2-1008w_k^5\left(w_k^7\right)^2-1008w_k^6\left(w_k^7\right)^2\\
&-668\left(w_k^3\right)^2w_k^4-886\left(w_k^3\right)^2w_k^5-960\left(w_k^4\right)^2w_k^5+2832w_k^3w_k^4w_k^5-768\left(w_k^3\right)^2w_k^6-960\left(w_k^4\right)^2w_k^6\\
&-3348\left(w_k^5\right)^2w_k^6+2160w_k^3w_k^4w_k^6+3552w_k^3w_k^5w_k^6-3696w_k^4w_k^5w_k^6+1348\left(w_k^3\right)^2w_k^7+2904\left(w_k^5\right)^2w_k^7\\
&+2088\left(w_k^6\right)^2w_k^7-1952w_k^3w_k^4w_k^7-4280w_k^3w_k^5w_k^7+2016w_k^4w_k^5w_k^7-3528w_k^3w_k^6w_k^7+2016w_k^4w_k^6w_k^7\\
&+5040w_k^5w_k^6w_k^7+36\left(w_k^2\right)^2\left(2w_k^4+w_k^5+6(w_k^7-4w_k^8)\right)-48(26\left(w_k^3\right)^2-54(w_k^5+w_k^6)w_k^3+27(w_k^5\\
&+w_k^6)^2)w_k^8+3w_k^2(11\left(w_k^3\right)^2+4(32w_k^4+37w_k^5+39w_k^6-126w_k^7+192w_k^8)w_k^3-336\left(w_k^7\right)^2-4(64\\
&\times\left(w_k^4\right)^2+140w_k^5w_k^4+96w_k^6w_k^4+87\left(w_k^5\right)^2+63\left(w_k^6\right)^2+138w_k^5w_k^6)+16(38w_k^4+65w_k^5+51w_k^6)w_k^7\\
&-864(w_k^5+w_k^6)w_k^8)\bigg]c_2^2-2c_3\left(324\left(w_k^1\right)^3-9(21w_k^2-67w_k^3+96w_k^4+145w_k^5+105w_k^6-104w_k^7)\left(w_k^1\right)^2\right.\\
&+9(41\left(w_k^3\right)^2-2(50w_k^4+82w_k^5+61w_k^6)w_k^3+w_k^2(-29w_k^3+84w_k^4+97w_k^5+63w_k^6)+(w_k^5+w_k^6)(108w_k^4\\
&+127w_k^5+81w_k^6))w_k^1-48(18w_k^2-20w_k^3+21(w_k^5+w_k^6))w_k^7w_k^1+74\left(w_k^3\right)^3-270\left(w_k^5\right)^3-162\\
&\times\left(w_k^6\right)^3+615w_k^3\left(w_k^5\right)^2-252w_k^4\left(w_k^5\right)^2+405w_k^3\left(w_k^6\right)^2-252w_k^4\left(w_k^6\right)^2-594w_k^5\left(w_k^6\right)^2-232\left(w_k^3\right)^2w_k^4\\
&-413\left(w_k^3\right)^2w_k^5+492w_k^3w_k^4w_k^5-315\left(w_k^3\right)^2w_k^6-702\left(w_k^5\right)^2w_k^6+492w_k^3w_k^4w_k^6+1020w_k^3w_k^5w_k^6\\
&-504w_k^4w_k^5w_k^6-3w_k^2(30\left(w_k^3\right)^2-(136w_k^4+173w_k^5+117w_k^6-152w_k^7)w_k^3+6(w_k^5+w_k^6)(26w_k^4+26w_k^5\\
&+15w_k^6-28w_k^7))-18\left(w_k^2\right)^2(8w_k^4+6w_k^5+3w_k^6-10w_k^7)+4(61\left(w_k^3\right)^2-126(w_k^5+w_k^6)w_k^3+63(w_k^5\\
&+w_k^6)^2)w_k^7\bigg)c_2+3c_3^2(2w_k^1-w_k^2+w_k^3-w_k^5-w_k^6)(3w_k^1+2w_k^3-3(w_k^5+w_k^6))(9w_k^1-6w_k^2+4w_k^3-3\\
&\times(w_k^5+w_k^6))\bigg]c_1+4c_2\bigg[27c_3^2(-2w_k^1+w_k^2-w_k^3+w_k^5+w_k^6)^3+9c_2c_3(21w_k^1+15w_k^3-4(7w_k^4+9w_k^5\\
&+6w_k^6-7w_k^7))(-2w_k^1+w_k^2-w_k^3+w_k^5+w_k^6)^2+c_2^2\bigg(-270\left(w_k^1\right)^3+9(9w_k^2-69w_k^3+128w_k^4+189w_k^5\\
&+165w_k^6-224w_k^7+192w_k^8)\left(w_k^1\right)^2+3\left(9\left(w_k^2\right)^2+6(11w_k^3-36w_k^4-50w_k^5-48w_k^6+84w_k^7-96w_k^8)\right.\\
&\times w_k^2+2\left(-74\left(w_k^3\right)^2+(236w_k^4+361w_k^5+309w_k^6-380w_k^7+288w_k^8)w_k^3-2(56\left(w_k^4\right)^2+202w_k^5w_k^4\right.\\
&+162w_k^6w_k^4+175\left(w_k^5\right)^2+126\left(w_k^6\right)^2+56\left(w_k^7\right)^2+294w_k^5w_k^6-2(56w_k^4+137w_k^5+117w_k^6)w_k^7)\\
&-288(w_k^5+w_k^6)w_k^8))w_k^1-101\left(w_k^3\right)^3+656\left(w_k^5\right)^3+432\left(w_k^6\right)^3-336w_k^3\left(w_k^4\right)^2-1224w_k^3\left(w_k^5\right)^2\\
&+920w_k^4\left(w_k^5\right)^2-900w_k^3\left(w_k^6\right)^2+648w_k^4\left(w_k^6\right)^2+1440w_k^5\left(w_k^6\right)^2-336w_k^3\left(w_k^7\right)^2+336w_k^5\left(w_k^7\right)^2\\
&+336w_k^6\left(w_k^7\right)^2+416\left(w_k^3\right)^2w_k^4+669\left(w_k^3\right)^2w_k^5+336\left(w_k^4\right)^2w_k^5-1336w_k^3w_k^4w_k^5+573\left(w_k^3\right)^2w_k^6\\
&+336\left(w_k^4\right)^2w_k^6+1680\left(w_k^5\right)^2w_k^6-1080w_k^3w_k^4w_k^6-2064w_k^3w_k^5w_k^6+1584w_k^4w_k^5w_k^6-632\left(w_k^3\right)^2w_k^7\\
&-1136\left(w_k^5\right)^2w_k^7-864\left(w_k^6\right)^2w_k^7+672w_k^3w_k^4w_k^7+1768w_k^3w_k^5w_k^7-672w_k^4w_k^5w_k^7+1512w_k^3w_k^6w_k^7\\
&-672w_k^4w_k^6w_k^7-2016w_k^5w_k^6w_k^7+432(-w_k^3+w_k^5+w_k^6)^2w_k^8+9\left(w_k^2\right)^2(w_k^3+4(w_k^4+2w_k^5+3w_k^6-7w_k^7\\
&+12w_k^8))+3w_k^2(31\left(w_k^3\right)^2-4(38w_k^4+53w_k^5+48w_k^6-74w_k^7+72w_k^8)w_k^3+4\left(28\left(w_k^4\right)^2+(80w_k^5+60w_k^6\right.\\
&-56w_k^7)w_k^4+61\left(w_k^5\right)^2+45\left(w_k^6\right)^2+28\left(w_k^7\right)^2+102w_k^5w_k^6-116w_k^5w_k^7-96w_k^6w_k^7+72(w_k^5+w_k^6)w_k^8)))\bigg]\bigg\},
\end{aligned}
$$

$$
w_{k+1}^8=\frac{2w_k^8}{C^{3/2}}+\frac{8}{27C^{9/2}}\bigg\{\bigg[112\left(w_k^4\right)^3+36w_k^2\left(w_k^4\right)^2+132w_k^5\left(w_k^4\right)^2-36w_k^6\left(w_k^4\right)^2-336w_k^7\left(w_k^4\right)^2 \tag{S12}
$$
$$
+78\left(w_k^5\right)^2w_k^4+54\left(w_k^6\right)^2w_k^4+336\left(w_k^7\right)^2w_k^4+108w_k^1w_k^6w_k^4-36w_k^2w_k^6w_k^4+216w_k^1w_k^7w_k^4+72w_k^2w_k^7w_k^4
$$

$$-624w_k^5 w_k^7 w_k^4 - 288 w_k^6 w_k^7 w_k^4 + 27\left(w_k^5\right)^3 + 27\left(w_k^6\right)^3 - 112\left(w_k^7\right)^3 + 81 w_k^5 \left(w_k^6\right)^2 - 468 w_k^1 \left(w_k^7\right)^2$$
$$+108 w_k^2 \left(w_k^7\right)^2 + 528 w_k^5 \left(w_k^7\right)^2 + 360 w_k^6 \left(w_k^7\right)^2 + 27\left(w_k^5\right)^2 w_k^6 + 54 w_k^1 w_k^5 w_k^6 - 348\left(w_k^5\right)^2 w_k^7 - 216 \left(w_k^6\right)^2$$
$$\times w_k^7 + 216 w_k^1 w_k^5 w_k^7 + 36 w_k^2 w_k^5 w_k^7 + 108 w_k^1 w_k^6 w_k^7 - 432 w_k^5 w_k^6 w_k^7 + 108 \big(3\left(w_k^1\right)^2 - 2(8 w_k^4 + 7 w_k^5 + 3 w_k^6$$
$$- 8 w_k^7) w_k^1 + (w_k^5 + w_k^6)(8 w_k^4 + 7 w_k^5 + 3 w_k^6 - 8 w_k^7) + 4 w_k^2 (2 w_k^4 + w_k^5 - 2 w_k^7)) w_k^8 + 6 \left(w_k^3\right)^2 (w_k^4 + 2 w_k^7$$
$$+ 24 w_k^8) + 6 w_k^3 \big( 2\left(w_k^4\right)^2 + 8 w_k^5 w_k^4 + 6 w_k^6 w_k^4 + 28 w_k^7 w_k^4 + 2\left(w_k^5\right)^2 - 46\left(w_k^7\right)^2 + 3 w_k^5 w_k^6 + 6 w_k^1 w_k^7$$
$$+ 24 w_k^5 w_k^7 + 18 w_k^6 w_k^7 + 72(w_k^1 - 2 w_k^4 - 2 w_k^5 - w_k^6 + 2 w_k^7) w_k^8 \big) \big] c_1^3 + 3 c_3 \big[ 9(10 w_k^4 + 7 w_k^5 + 3 w_k^6 - 24 w_k^7$$
$$+ 48 w_k^8) \left(w_k^1\right)^2 + 6 \big( - 48\left(w_k^7\right)^2 + 12(w_k^2 + 8 w_k^4 + 9 w_k^5 + 5 w_k^6) w_k^7 - 3(16\left(w_k^4\right)^2 + 26 w_k^5 w_k^4 + 11\left(w_k^5\right)^2$$
$$+ 3\left(w_k^6\right)^2 + 10(w_k^4 + w_k^5) w_k^6) - 72(w_k^2 + w_k^5 + w_k^6) w_k^8 + 2 w_k^3 (10 w_k^4 + 7 w_k^5 + 3 w_k^6 - 22 w_k^7 + 36 w_k^8)) w_k^1$$
$$+ 4\left(w_k^3\right)^2 (10 w_k^4 + 7 w_k^5 + 3 w_k^6 - 20 w_k^7 + 27 w_k^8) + 12 w_k^3 \big( - 12\left(w_k^4\right)^2 - 2(11 w_k^5 + 5 w_k^6 - 12 w_k^7) w_k^4$$
$$- 10\left(w_k^5\right)^2 - 3\left(w_k^6\right)^2 - 12\left(w_k^7\right)^2 - 10 w_k^5 w_k^6 + 4 w_k^2 w_k^7 + 30 w_k^5 w_k^7 + 18 w_k^6 w_k^7 - 18(w_k^2 + w_k^5 + w_k^6) w_k^8 \big)$$
$$+ 9(12 w_k^8 \left(w_k^2\right)^2 + 4(4\left(w_k^4\right)^2 + 4 w_k^5 w_k^4 - 8 w_k^7 w_k^4 + \left(w_k^5\right)^2 + 4\left(w_k^7\right)^2 - 6 w_k^5 w_k^7 - 2 w_k^6 w_k^7 + 6(w_k^5$$
$$+ w_k^6) w_k^8) w_k^2 + (w_k^5 + w_k^6)(16\left(w_k^4\right)^2 + 26 w_k^5 w_k^4 + 10 w_k^6 w_k^4 - 32 w_k^7 w_k^4 + 11\left(w_k^5\right)^2 + 3\left(w_k^6\right)^2 + 16\left(w_k^7\right)^2$$
$$+ 10 w_k^5 w_k^6 - 32 w_k^5 w_k^7 - 16 w_k^6 w_k^7 + 12(w_k^5 + w_k^6) w_k^8)) \big] c_1^2 - 9 c_3^2 (2 w_k^1 - w_k^2 + w_k^3 - w_k^5 - w_k^6) \big[ 9\left(w_k^1\right)^2$$
$$+ 6(2 w_k^3 - 8 w_k^4 - 7 w_k^5 - 3 w_k^6 + 8 w_k^7) w_k^1 + 4\left(w_k^3\right)^2 + 3(w_k^5 + w_k^6)(8 w_k^4 + 7 w_k^5 + 3 w_k^6 - 8 w_k^7)$$
$$+ 12 w_k^2 (2 w_k^4 + w_k^5 - 2 w_k^7) - 12 w_k^3 (2 w_k^4 + 2 w_k^5 + w_k^6 - 2 w_k^7) \big] c_1 + 27 c_3^3 (-2 w_k^1 + w_k^2 - w_k^3 + w_k^5 + w_k^6)^3$$
$$+ 2 c_2^3 \big[ 27\left(w_k^1\right)^3 + 27(3 w_k^2 + 7 w_k^3 - 2(22 w_k^4 + 32 w_k^5 + 33 w_k^6 - 70 w_k^7 + 96 w_k^8)) \left(w_k^1\right)^2 + 9 (21\left(w_k^3\right)^2$$
$$- 4(46 w_k^4 + 66 w_k^5 + 63 w_k^6 - 130 w_k^7 + 168 w_k^8) w_k^3 + 16(13\left(w_k^4\right)^2 + 37 w_k^5 w_k^4 + 30 w_k^6 w_k^4 - 50 w_k^7 w_k^4$$
$$+ 27\left(w_k^5\right)^2 + 18\left(w_k^6\right)^2 + 37\left(w_k^7\right)^2 + 45 w_k^5 w_k^6 - 76 w_k^5 w_k^7 - 57 w_k^6 w_k^7 + 6(8 w_k^4 + 13 w_k^5 + 9 w_k^6 - 8 w_k^7) w_k^8)$$
$$+ 6 w_k^2 (w_k^3 + 4(w_k^4 + 2 w_k^5 + 3 w_k^6 - 7 w_k^7 + 12 w_k^8))) w_k^1 + 51\left(w_k^3\right)^3 - 448\left(w_k^4\right)^3 - 1824\left(w_k^5\right)^3 - 864\left(w_k^6\right)^3$$
$$+ 448\left(w_k^7\right)^3 + 1104 w_k^3 \left(w_k^4\right)^2 + 2376 w_k^3 \left(w_k^5\right)^2 - 3408 w_k^4 \left(w_k^5\right)^2 + 1512 w_k^3 \left(w_k^6\right)^2 - 2160 w_k^4 \left(w_k^6\right)^2$$
$$- 3456 w_k^5 \left(w_k^6\right)^2 + 2832 w_k^3 \left(w_k^7\right)^2 - 1344 w_k^4 \left(w_k^7\right)^2 - 3840 w_k^5 \left(w_k^7\right)^2 - 3168 w_k^6 \left(w_k^7\right)^2 - 564\left(w_k^3\right)^2 w_k^4$$
$$- 792\left(w_k^3\right)^2 w_k^5 - 2112\left(w_k^4\right)^2 w_k^5 + 3216 w_k^3 w_k^4 w_k^5 - 702\left(w_k^3\right)^2 w_k^6 - 1440\left(w_k^4\right)^2 w_k^6 - 4320\left(w_k^5\right)^2 w_k^6$$
$$+ 2592 w_k^3 w_k^4 w_k^6 + 3888 w_k^3 w_k^5 w_k^6 - 5184 w_k^4 w_k^5 w_k^6 + 1428\left(w_k^3\right)^2 w_k^7 + 1344\left(w_k^4\right)^2 w_k^7 + 5568\left(w_k^5\right)^2 w_k^7$$
$$+ 3456\left(w_k^6\right)^2 w_k^7 - 3936 w_k^3 w_k^4 w_k^7 - 6240 w_k^3 w_k^5 w_k^7 + 5952 w_k^4 w_k^5 w_k^7 - 4752 w_k^3 w_k^6 w_k^7 + 4608 w_k^4 w_k^6 w_k^7$$
$$+ 8640 w_k^5 w_k^6 w_k^7 - 864(w_k^3 - w_k^5 - w_k^6)(2 w_k^3 - 4 w_k^4 - 5 w_k^5 - 3 w_k^6 + 4 w_k^7) w_k^8 + 9 w_k^2 \big(\left(w_k^3\right)^2 + 8(3 w_k^4$$
$$+ 5 w_k^5 + 6 w_k^6 - 15 w_k^7 + 24 w_k^8) w_k^3 - 240\left(w_k^7\right)^2 - 8(6\left(w_k^4\right)^2 + 18(w_k^5 + w_k^6) w_k^4 + 13\left(w_k^5\right)^2 + 9\left(w_k^6\right)^2 + 24 w_k^5$$
$$\times w_k^6) + 96(3 w_k^4 + 4 w_k^5 + 3 w_k^6) w_k^7 - 96(4 w_k^4 + 5 w_k^5 + 3 w_k^6 - 4 w_k^7) w_k^8) \big] + 12 c_2^2 \big[ c_1 \big( \left(w_k^3\right)^3 + (34 w_k^4 + 36 w_k^5$$
$$+ 33 w_k^6 - 170 w_k^7 + 336 w_k^8) \left(w_k^3\right)^2 + \big( - 160\left(w_k^4\right)^2 - 364 w_k^5 w_k^4 - 288 w_k^6 w_k^4 + 712 w_k^7 w_k^4 - 218\left(w_k^5\right)^2$$
$$- 135\left(w_k^6\right)^2 - 564\left(w_k^7\right)^2 - 348 w_k^5 w_k^6 + 944 w_k^5 w_k^7 + 684 w_k^6 w_k^7 + 6 w_k^2 (2 w_k^4 + w_k^5 + 14 w_k^7 - 48 w_k^8) - 432$$
$$\times (2 w_k^4 + 3 w_k^5 + 2 w_k^6 - 2 w_k^7) w_k^8) w_k^3 + 112\left(w_k^4\right)^3 + 232\left(w_k^5\right)^3 + 108\left(w_k^6\right)^3 - 112\left(w_k^7\right)^3 + 24 w_k^2 \left(w_k^4\right)^2$$
$$+ 42 w_k^2 \left(w_k^5\right)^2 + 510 w_k^4 \left(w_k^5\right)^2 + 27 w_k^2 \left(w_k^6\right)^2 + 306 w_k^4 \left(w_k^6\right)^2 + 432 w_k^5 \left(w_k^6\right)^2 + 396 w_k^2 \left(w_k^7\right)^2 + 336 w_k^4 \left(w_k^7\right)^2$$
$$+ 816 w_k^5 \left(w_k^7\right)^2 + 648 w_k^6 \left(w_k^7\right)^2 + 396\left(w_k^4\right)^2 w_k^5 + 72 w_k^2 w_k^4 w_k^5 + 228\left(w_k^4\right)^2 w_k^6 + 522\left(w_k^5\right)^2 w_k^6 + 108 w_k^2$$
$$\times w_k^4 w_k^6 + 90 w_k^2 w_k^5 w_k^6 + 708 w_k^4 w_k^5 w_k^6 - 336\left(w_k^4\right)^2 w_k^7 - 972\left(w_k^5\right)^2 w_k^7 - 576\left(w_k^6\right)^2 w_k^7 - 408 w_k^2 w_k^4 w_k^7 - 420$$
$$\times w_k^2 w_k^5 w_k^7 - 1200 w_k^4 w_k^5 w_k^7 - 288 w_k^2 w_k^6 w_k^7 - 864 w_k^4 w_k^6 w_k^7 - 1440 w_k^5 w_k^6 w_k^7 + 108((w_k^5 + w_k^6)(8 w_k^4 + 9 w_k^5$$
$$+ 5 w_k^6 - 8 w_k^7) + 4 w_k^2 (2 w_k^4 + 2 w_k^5 + w_k^6 - 2 w_k^7)) w_k^8 + 9\left(w_k^1\right)^2 (w_k^3 + 4 w_k^4 + 7 w_k^5 + 6 w_k^6 - 46 w_k^7 + 108 w_k^8)$$
$$+ 3 w_k^1 \big( 2\left(w_k^3\right)^2 + 2(12 w_k^4 + 15 w_k^5 + 15 w_k^6 - 88 w_k^7 + 192 w_k^8) w_k^3 - 84\left(w_k^4\right)^2 - 113\left(w_k^5\right)^2 - 72\left(w_k^6\right)^2$$
$$- 348\left(w_k^7\right)^2 - 188 w_k^4 w_k^5 - 144 w_k^4 w_k^6 - 174 w_k^5 w_k^6 + 424 w_k^4 w_k^7 + 536 w_k^5 w_k^7 + 372 w_k^6 w_k^7 + 6 w_k^2 (2 w_k^4 + w_k^5$$
$$+ 6 w_k^7 - 24 w_k^8) - 72(8 w_k^4 + 11 w_k^5 + 7 w_k^6 - 8 w_k^7) w_k^8 \big) \big) - 9 c_3 (2 w_k^1 - w_k^2 + w_k^3 - w_k^5 - w_k^6)(9\left(w_k^1\right)^2 + 4(3 w_k^3$$
$$- 7 w_k^4 - 9 w_k^5 - 6 w_k^6 + 10 w_k^7 - 6 w_k^8) w_k^1 + 4\left(w_k^3\right)^2 + 16\left(w_k^4\right)^2 + 28\left(w_k^5\right)^2 + 12\left(w_k^6\right)^2 + 16\left(w_k^7\right)^2 + 2 w_k^2 w_k^4$$
$$+ 3 w_k^2 w_k^5 + 42 w_k^4 w_k^5 + 3 w_k^2 w_k^6 + 26 w_k^4 w_k^6 + 36 w_k^5 w_k^6 - 8 w_k^2 w_k^7 - 32 w_k^4 w_k^7 - 48 w_k^5 w_k^7 - 32 w_k^6 w_k^7 + 12(w_k^2$$
$$+ w_k^5 + w_k^6) w_k^8 - w_k^3 (18 w_k^4 + 23 w_k^5 + 15 w_k^6 - 24 w_k^7 + 12 w_k^8)) \big] + 3 c_2 \big[ (9(3 w_k^6 + 22 w_k^7 - 144 w_k^8) \left(w_k^1\right)^2$$

$$
\begin{aligned}
&+6\big(20\left(w_k^4\right)^2 + 40w_k^5 w_k^4 + 12w_k^6 w_k^4 - 248w_k^7 w_k^4 + 23\left(w_k^5\right)^2 + 18\left(w_k^6\right)^2 + 252\left(w_k^7\right)^2 + 24w_k^5 w_k^6 + 6w_k^2 \\
&\times w_k^7 - 256w_k^5 w_k^7 - 156w_k^6 w_k^7 + 3w_k^3(4w_k^4 + 2w_k^5 + w_k^6 + 14w_k^7 - 88w_k^8) + 72(w_k^2 + 8w_k^4 + 9w_k^5 + 5w_k^6 \\
&-8w_k^7)w_k^8)w_k^1 + \left(w_k^3\right)^2(24w_k^4 + 12w_k^5 + 3w_k^6 + 94w_k^7 - 480w_k^8) + 2w_k^3(52\left(w_k^4\right)^2 + (80w_k^5 + 60w_k^6 \\
&-440w_k^7 + 864w_k^8)w_k^4 + 43\left(w_k^5\right)^2 + 8w_k^5(9w_k^6 - 59w_k^7 + 135w_k^8) + 6(6\left(w_k^6\right)^2 - 54(w_k^7 - 2w_k^8)w_k^6 + w_k^7(w_k^2 \\
&+70w_k^7) + 24(w_k^2 - 6w_k^7)w_k^8)) + 2(112\left(w_k^7\right)^3 - 168(2w_k^4 + 4w_k^5 + 3w_k^6)\left(w_k^7\right)^2 + 24(14\left(w_k^4\right)^2 + 38w_k^5 w_k^4 \\
&+24w_k^6 w_k^4 + 26\left(w_k^5\right)^2 + 15\left(w_k^6\right)^2 + 36w_k^5 w_k^6)w_k^7 - 2(56\left(w_k^4\right)^3 + 12(11w_k^5 + 4w_k^6)\left(w_k^4\right)^2 + 6(21\left(w_k^5\right)^2 \\
&+22w_k^6 w_k^5 + 12\left(w_k^6\right)^2)w_k^4 + 49\left(w_k^5\right)^3 + 27\left(w_k^6\right)^3 + 99w_k^5\left(w_k^6\right)^2 + 99\left(w_k^5\right)^2 w_k^6) - 432(w_k^5 + w_k^6)(2w_k^4 \\
&+2w_k^5 + w_k^6 - 2w_k^7)w_k^8 + 3w_k^2(12\left(w_k^4\right)^2 + 12w_k^5 w_k^4 + 56w_k^7 w_k^4 + 3\left(w_k^5\right)^2 - 84\left(w_k^7\right)^2 + 40w_k^5 w_k^7 + 24w_k^6 w_k^7 \\
&-72(4w_k^4 + 3w_k^5 + w_k^6 - 4w_k^7)w_k^8)))c_1^2 + 4c_3\big(27\left(w_k^1\right)^3 + 9(6w_k^3 - 32w_k^4 - 29w_k^5 - 15w_k^6 + 52w_k^7 - 48w_k^8) \\
&\times\left(w_k^1\right)^2 + 3(12\left(w_k^3\right)^2 - (114w_k^4 + 107w_k^5 + 57w_k^6 - 176w_k^7 + 144w_k^8)w_k^3 + 3(32\left(w_k^4\right)^2 + 14w_k^2 w_k^4 \\
&+82w_k^5 w_k^4 + 50w_k^6 w_k^4 - 64w_k^7 w_k^4 + 48\left(w_k^5\right)^2 + 18\left(w_k^6\right)^2 + 32\left(w_k^7\right)^2 + 9w_k^2 w_k^5 + 3w_k^2 w_k^6 + 58w_k^5 w_k^6 \\
&-32w_k^2 w_k^7 - 104w_k^5 w_k^7 - 72w_k^6 w_k^7 + 48(w_k^2 + w_k^5 + w_k^6)w_k^8))w_k^1 + 8\left(w_k^3\right)^3 - 2\left(w_k^3\right)^2(50w_k^4 + 49w_k^5 + 27w_k^6 \\
&-74w_k^7 + 54w_k^8) + 3w_k^3(48\left(w_k^4\right)^2 + 28w_k^2 w_k^4 + 134w_k^5 w_k^4 + 86w_k^6 w_k^4 - 96w_k^7 w_k^4 + 83\left(w_k^5\right)^2 + 33\left(w_k^6\right)^2 \\
&+48\left(w_k^7\right)^2 + 18w_k^2 w_k^5 + 6w_k^2 w_k^6 + 104w_k^5 w_k^6 - 56w_k^2 w_k^7 - 168w_k^5 w_k^7 - 120w_k^6 w_k^7 + 72(w_k^2 + w_k^5 + w_k^6)w_k^8) \\
&+9(4(w_k^7 - 3w_k^8)\left(w_k^2\right)^2 - (16\left(w_k^4\right)^2 + 30w_k^5 w_k^4 + 14w_k^6 w_k^4 - 32w_k^7 w_k^4 + 13\left(w_k^5\right)^2 + 3\left(w_k^6\right)^2 + 16\left(w_k^7\right)^2 \\
&+12w_k^5 w_k^6 - 40w_k^5 w_k^7 - 24w_k^6 w_k^7 + 24(w_k^5 + w_k^6)w_k^8)w_k^2 - 2(w_k^5 + w_k^6)(8\left(w_k^4\right)^2 + 17w_k^5 w_k^4 + 9w_k^6 w_k^4 \\
&-16w_k^7 w_k^4 + 9\left(w_k^5\right)^2 + 3\left(w_k^6\right)^2 + 8\left(w_k^7\right)^2 + 10w_k^5 w_k^6 - 20w_k^5 w_k^7 - 12w_k^6 w_k^7 + 6(w_k^5 + w_k^6)w_k^8)))c_1 + 36 \\
&\times c_3^2(-2w_k^1 + w_k^2 - w_k^3 + w_k^5 + w_k^6)^2(3w_k^1 + 2w_k^3 - 4w_k^4 - 5w_k^5 - 3w_k^6 + 4w_k^7)\big]\big\}.
\end{aligned}
$$

It is straightforward to show that in the zero-field limit $s_k = r_k^2 = r_k^3 = w_k^2 = w_k^3 = \cdots = w_k^8 = 0$, and Eqs. (S2), (S5) reduce to the RG equations for the hierarchical Edwards-Anderson model [2].

## S2. RENORMALIZATION-GROUP FLOW IN TERMS OF PHYSICAL PARAMETERS OF THE MODEL

Here we will discuss how to express the RG flow of the coefficients

$$s_k, \{r_k^p\}, \{w_k^p\} \tag{S13}$$

in terms of the physical parameters of the model, i.e. the interaction-decay exponent $\sigma$, the inverse temperature $\beta$ and the magnetic-field strength $\varsigma_h^2 \equiv \mathbb{E}[h_i^2]$. While the RG FPs can be completely determined by imposing that the FP is invariant with respect to the RG transformation, to determine the RG coefficients for any finite $k$ an explicit solution of the RG recurrence Eqs. (S1)-(S12) is required. Given that the solution of these equations is particularly involved, we will not provide the explicit expression of $s_k, \{r_k^p\}, \{w_k^p\}$ as functions of $\sigma$, $\beta$, $\varsigma_h$ here, but we will rather lay out the procedure to determine it. First, we will relate the coefficients (S13) to the averages of moments of the overlap $Q_{ab}$—see for example the RHS of Eq. (19). Second, we will write the average moments of $Q_{ab}$ for $k = 0$ in terms of $\sigma$, $\beta$ and $\varsigma_h$: as a result, we will obtain an implicit expression of $s_0, \{r_0^p\}, \{w_0^p\}$ as functions of the physical parameters of the model. The dependence of the RG coefficients (S13) on $\sigma$, $\beta$ and $\varsigma_h$ can then be obtained implicitly from $s_0, \{r_0^p\}, \{w_0^p\}$ by iterating the RG Eqs. (S1)-(S12).

In order to illustrate this procedure, let us introduce the indexes $A, B, C, \ldots$, which run over replica pairs $(1,2), (1,3), \ldots, (n-1,n)$. In addition, let us introduce the coupling constants

$$g_{ABC}^p \equiv \frac{w_k^p}{6} \frac{\partial^3 \mathrm{I}_p^3[Q]}{\partial Q_A \partial Q_B \partial Q_C}. \tag{S14}$$

By using Eqs. (7), (S14), the replicated partition function can be rewritten as

$$\mathcal{Z}_k[Q] = \exp\left(-2s_k \sum_A Q_A - \frac{C}{4}\sum_{AB} Q_A \mathscr{F}_{AB} Q_B - \frac{1}{6}\sum_{p=1}^8 g_{ABC}^p Q_A Q_B Q_C\right). \tag{S15}$$

Equation (S15) has the structure of a field theory with a quadratic term and a non-quadratic correction: this form allows for extracting easily the overlap moments by means of diagrammatic methods [3]. The result is

$$\overline{Q_A} = -\frac{4s_k\mathscr{H}}{C} - \frac{2}{C^2}\sum_{p=1}^{8}\sum_{B,C,D} g^p_{BCD}\mathscr{F}^{-1}_{AB}\left[\frac{2}{C}(2s_k\mathscr{H})^2 + \mathscr{F}^{-1}_{CD}\right] + O(g^2), \qquad (S16)$$

$$\overline{Q_AQ_B} = \left(\frac{4s_k\mathscr{H}}{C}\right)^2 + \frac{2}{C}\mathscr{F}^{-1}_{AB} + \frac{16s_k\mathscr{H}}{C^3}\sum_{p=1}^{8} g^p_{CDE}\mathscr{F}^{-1}_{AC}\mathscr{F}^{-1}_{BD} + O(g^2), \qquad (S17)$$

$$\overline{Q_AQ_BQ_C} = -\left(\frac{4s_k\mathscr{H}}{C}\right)^3 - \frac{8s_k\mathscr{H}}{C^2}\left(\mathscr{F}^{-1}_{AB} + \mathscr{F}^{-1}_{AC} + \mathscr{F}^{-1}_{BC}\right) - \left(\frac{2}{C}\right)^3\sum_{p=1}^{8} g^p_{DEF}\,\mathscr{F}^{-1}_{AD}\mathscr{F}^{-1}_{BE}\mathscr{F}^{-1}_{CF} + O(g^2), \quad (S18)$$

where $\mathscr{H} \equiv \sum_{a<b=1}^{n}\mathscr{F}^{-1}_{12,ab}$

We will now write the LHSs of Eqs. (S16)-(S18) in terms of Boltzmann averages of overlap products $\langle q_{12}\rangle, \langle q_{12}^2\rangle$, ..., where in what follows the average $\langle\rangle$ of a function of multiple replicas will denote the average over all replicas, each replica having an independent Boltzmann measure with Hamiltonian (1). Let us start with Eq. (S16): we have

$$\mathbb{E}[\langle q_{12}\rangle] = \lim_{n\to 0}\int[dQ]Z_k[Q]Q_{12} \qquad (S19)$$
$$= \lim_{n\to 0} C^{-\left(\frac{n(n-1)}{2}+1\right)\frac{k}{2}}\int[dQ]\mathcal{Z}_k[Q]Q_{12},$$
$$= C^{-k/2}\lim_{n\to 0}\overline{Q_{12}},$$

where in the first line of Eq. (S19) we used Eq. (2), in the second line we used Eq. (5), and in the third line we utilized the normalization identity $\lim_{n\to 0}\int[dQ]\mathcal{Z}_k[Q] = 1$ to write the integral in the second line as an average with respect to $\mathcal{Z}_k[Q]$, which we denote by $\overline{\phantom{-}}$. Proceeding along the lines of Eq. (S19), we obtain

$$\mathbb{E}[\langle q_Aq_B\rangle] = C^{-k}\lim_{n\to 0}\overline{Q_AQ_B}, \qquad (S20)$$
$$\mathbb{E}[\langle q_Aq_Bq_C\rangle] = C^{-3k/2}\lim_{n\to 0}\overline{Q_AQ_BQ_C}. \qquad (S21)$$

For $k=0$, we can write the RHS of Eqs. (S19)-(S21) as functions to the physical parameters $\beta, \varsigma_h$:

$$\mathbb{E}[\langle q_{12}\rangle] = \mathbb{E}[\langle q_{12}q_{13}\rangle] = \mathbb{E}[\langle q_{12}^3\rangle] = \mathbb{E}[\langle q_{12}^2q_{13}\rangle] = \mathbb{E}[\langle q_{12}^2q_{34}\rangle] = \mathbb{E}[\langle q_{12}q_{13}q_{24}\rangle] = \mathbb{E}[\tanh(\beta h)^2], \qquad (S22)$$
$$\mathbb{E}[\langle q_{12}^2\rangle] = \mathbb{E}[\langle q_{12}q_{23}q_{31}\rangle] = 1, \qquad (S23)$$
$$\mathbb{E}[\langle q_{12}q_{34}\rangle] = \mathbb{E}[\langle q_{12}q_{13}q_{45}\rangle] = \mathbb{E}[\tanh(\beta h)^4], \qquad (S24)$$
$$\mathbb{E}[\langle q_{12}q_{13}q_{14}\rangle] = 0, \qquad (S25)$$
$$\mathbb{E}[\langle q_{12}q_{34}q_{56}\rangle] = \mathbb{E}[\tanh(\beta h)^6], \qquad (S26)$$

where the expectation values of powers of $\tanh(h)$ in Eqs. (S22), (S24), (S26) depend on $\beta, \varsigma_h$ only through the product $\beta\varsigma_h$.

By using Eqs. (S16)-(S18), (S19)-(S21), (S22)-(S26) we obtain a set of twelve equations which relate the RG coefficients (S13) for $k=0$ to $\sigma$ and $\beta\varsigma_h$:

$$s_0 = s_0(\sigma, \beta\varsigma_h),\ r_0^p = r_0^p(\sigma, \beta\varsigma_h),\ w_0^p = w_0^p(\sigma, \beta\varsigma_h), \qquad (S27)$$

where in the LHSs of Eqs. (S27) we imply the $n\to 0$ limit. The RG coefficients for any value of $k$ can then be related implicitly to $\sigma$ and $\beta\varsigma_h$ by iterating of the recursive Eqs. (S1)-(S12) with initial condition (S27).

---



\end{document}